%% file: main.tex
\documentclass[12pt,a4paper]{elsarticle}

\usepackage[colorlinks,citecolor=blue,linktoc=all,linkcolor=cyan]{hyperref}
\usepackage{graphicx}

\usepackage[T1]{fontenc}
\usepackage{mathrsfs}             
\usepackage{slashed}              
\usepackage{amsmath}
\usepackage{amssymb}
\usepackage{amsbsy}  
\usepackage{amsfonts} 

\usepackage{wasysym} 
\usepackage{booktabs} 

\numberwithin{equation}{section}
\numberwithin{table}{section}
\numberwithin{figure}{section}

\usepackage{placeins}         
\usepackage{amsmath, amssymb} 
\usepackage{geometry}
\usepackage{graphicx}
\usepackage{hyperref}
\usepackage{booktabs}
\usepackage{siunitx}
\renewcommand{\tablename}{Table}

\begin{document}
\begin{frontmatter}

\title{
Worldwide Reactor Neutrino Propagation to Underground Labs: Matter Effects and Flux Predictions
}
\author[1,2]{Keyu Han}
\author[3]{Juncheng Qian}
\author[1,2]{Shaomin Chen}
\address[1]{Center for high energy physics, Tsinghua University, Beijing 100084, China}
\address[2]{Department of Engineering Physics, Tsinghua University, Beijing 10084, China}
\address[3]{Tsinglan School, 8 Tainan Road, Songshan Lake, Dongguan 523808, Guangdong, China}

\begin{abstract}
As a unique probe for geophysical research, geoneutrinos can reveal the distribution of internal heat sources in the Earth by detecting electron antineutrinos produced by the radioactive decay of $^{238}$U, $^{232}$Th, and $^{40}$K. However, commercial nuclear power plants continuously produce the same type of electron antineutrinos, which constitute a primary background difficult to eliminate in geoneutrino experiments. As geoneutrino measurements and reactor background modeling approach sub-percent precision, even small matter-induced corrections to reactor antineutrino propagation require quantitative assessment.
In this paper, we develop a high-precision prediction framework for reactor neutrino fluxes at underground labs, using global reactor operating data, reactor-to-detector distances, and matter effects (MSW) on neutrino propagation through the Earth. 
To solve the three-flavor MSW evolution efficiently, we implement a second-order Strang-splitting solver in the vacuum mass basis. Within this framework, we have calculated the reactor neutrino oscillation probabilities, including the MSW effect under one-dimensional (spherically symmetric) and three-dimensional (including lateral inhomogeneities) Earth models, and compared them with the vacuum oscillation scenario, to assess the impact of Earth's structural features on the accuracy of reactor neutrino flux predictions.

\end{abstract}
\end{frontmatter}

\section{Introduction}
Following the successful validation of the standard solar model by solar neutrino observations, geoneutrino research has emerged as a prominent interdisciplinary field that bridges neutrino physics and Earth sciences~\cite{Sramek:2012hma, Ludhova:2013hna}. The primary objective of this research is to quantify the heat contribution from radioactive isotopes within the Earth's interior. 
To date, KamLAND and Borexino have observed $\bar{\nu}_e$'s signals from the decay chains of uranium-238 ($^{238}$U) and thorium-232 ($^{232}$Th) via inverse beta decay (IBD)~\cite{Araki:2005qa, Borexino:2019gps}. Unfortunately, because the energy of $\bar{\nu}_e$'s from potassium-40 ($^{40}$K) decay is below the IBD reaction threshold (1.806 MeV), detection via this method is not feasible. Other ongoing or planned neutrino experiments, such as JUNO, SNO+, and JNE~\cite{Chen:2005zza, Han:2015roa, Jinping:2016iiq, Wang:2017etb}, also have geoneutrino programs that aim not only to detect all three components but also to provide directional information, since comprehensive measurements can provide direct constraints on the Earth's energy budget and distribution and allow comparisons with predictions from geophysical models~\cite{Bellini:2021sow, McDonough:2019ldt, Dye:2011mc, Fiorentini:2007te, Sramek:2012nk}.

Unlike the study of reactor neutrinos, the detection of geoneutrinos is complicated by substantial background interference from commercial nuclear power plants, as both involve $\bar{\nu}_e$'s and have significant spectral overlap below approximately 3.3 MeV. From an experimental perspective, the farther the detector is from the nuclear power plant, the smaller the background interference will be; ideally, it should be as far away as possible. However, in reality, it is almost impossible to eliminate the reactor neutrino background. In decomposing the energy spectrum, the signals from geoneutrinos and reactor neutrinos overlap significantly, leaving the current measurement uncertainty at 20\% to 30\% and preventing higher precision, not to mention the determination of the U/Th ratio~\cite{Wan:2016nhe}. This limitation severely affects the precise quantification of radioactive heat production within the Earth and poses a significant challenge to the future verification of geophysical models.

According to the latest data from the International Atomic Energy Agency (IAEA), there are currently hundreds of nuclear reactors in operation or under construction worldwide. The neutrinos produced by these reactors are radiated isotropically, and their flux decreases in inverse proportion to the square of the distance from underground labs. To ensure measurement accuracy for geoneutrinos at the few-percent level, it is necessary to quantify the contributions from all reactors precisely. 

It is known that reactor $\bar{\nu}_e$'s propagating through the Earth experience matter effects through the Mikheyev--Smirnov--Wolfenstein (MSW) mechanism. In the MeV energy range, these corrections are usually smaller than 1\% and have therefore often been neglected in reactor-$\bar{\nu}_e$ studies. However, the relevant question is no longer whether the Earth-matter effect is large in an absolute sense, but whether it remains negligible for the level of precision now targeted in reactor-background modeling. As geoneutrino measurements and related reactor-background predictions move toward sub-percent precision, corrections at the $10^{-3}$--$10^{-2}$ level warrant a dedicated quantitative treatment rather than being discarded a priori.

A significant challenge in this context is computational complexity. A global reactor-background prediction requires repeated three-flavor propagation calculations across numerous reactor-detector trajectories, energy bins, Earth models, and Monte Carlo samples used for uncertainty propagation. Repeated diagonalization of the full Hamiltonian at every propagation step, therefore, becomes computationally intensive during production calculations. This situation underscores the need for an efficient and high-precision solver for the MSW evolution. In this work, we systematically compile publicly available data on the global reactor fleet, calculate precise reactor-detector baselines for underground laboratories, and evaluate reactor $\bar{\nu}_e$ oscillation probabilities in both one-dimensional (spherically symmetric) and three-dimensional (including lateral inhomogeneities) Earth models. As a baseline numerical treatment of matter propagation, we implement a second-order Strang-splitting approach and compare its predictions with those derived from vacuum conditions to quantify the MSW correction and its dependence on Earth-model variations in precision reactor-background predictions.

\input{sec2}

\input{sec3}

\input{sec4}

\input{sec5}

\input{sec6}

\input{sec7}

\section{Summary}
Geoneutrinos, an emerging probe of Earth's internal structure and energy sources, provide a unique window into the composition of the Earth's core and into radioactive decay processes. With the ongoing expansion of underground labs worldwide, a MeV-scale neutrino detection network has been preliminarily established, encompassing neutrino signals from the decay of naturally occurring radioactive isotopes in the Earth. However, the significant similarities between geoneutrinos and reactor neutrinos in their energy spectra and temporal structures pose a critical challenge for the precise quantification of the reactor neutrino flux and background suppression.

In this work, we constructed a high-precision prediction framework for reactor $\bar{\nu}_e$ fluxes and IBD event rates at underground laboratories using worldwide reactor operating data, reactor-to-detector baselines, and three-flavor neutrino propagation in Earth matter.
For the MSW treatment, we implemented a second-order Strang-splitting solver in the vacuum mass basis as the baseline numerical method for matter propagation. We compared it with a first-order analytical approximation. In the convergence test for CJPL, the Strang result becomes numerically indistinguishable from the reference solution obtained by direct step-by-step diagonalization at the production setting adopted in this work.

Using this framework, we evaluated reactor $\bar{\nu}_e$ predictions under vacuum, 1D PREM, and hybrid 3D Earth-density models. Our findings demonstrate that high-precision models that incorporate Earth-matter effects can improve the precision and robustness of reactor neutrino flux predictions, providing a theoretical foundation for background suppression in the global geoneutrino detection network. For CJPL, the inclusion of the MSW effect increases the integrated reactor $\bar{\nu}_e$ flux from $7.46\times10^4$ to $7.48\times10^4\ \mathrm{cm^{-2}\,s^{-1}}$, corresponding to a correction of about 0.3\%, while the IBD event rate increases from 24.14 to 24.30 TNU, corresponding to about 0.7\%. Across the underground laboratories considered in this work, the Earth-model dependence ranges from 0.05\% to 0.53\%, with the largest sensitivity observed at JUNO and Yemilab. These results show that although the Earth-matter effect remains modest in the MeV energy range, it already enters the sub-percent regime relevant for future precision reactor-background predictions and should be included in next-generation geoneutrino analyses.


\appendix
\input{Appendix} 

\section*{Acknowledgments}
This work is funded by the National Natural Science Foundation of China under Grant No. 12127808
and the S-plan of Tsinglan School.

\bibliography{mybibfile}

\end{document}

%% file: sec2.tex
\section{Worldwide nuclear reactor survey}
\label{sec:reactor_survey}
\begin{figure}
    \centering
    \includegraphics[width=1.0\linewidth]{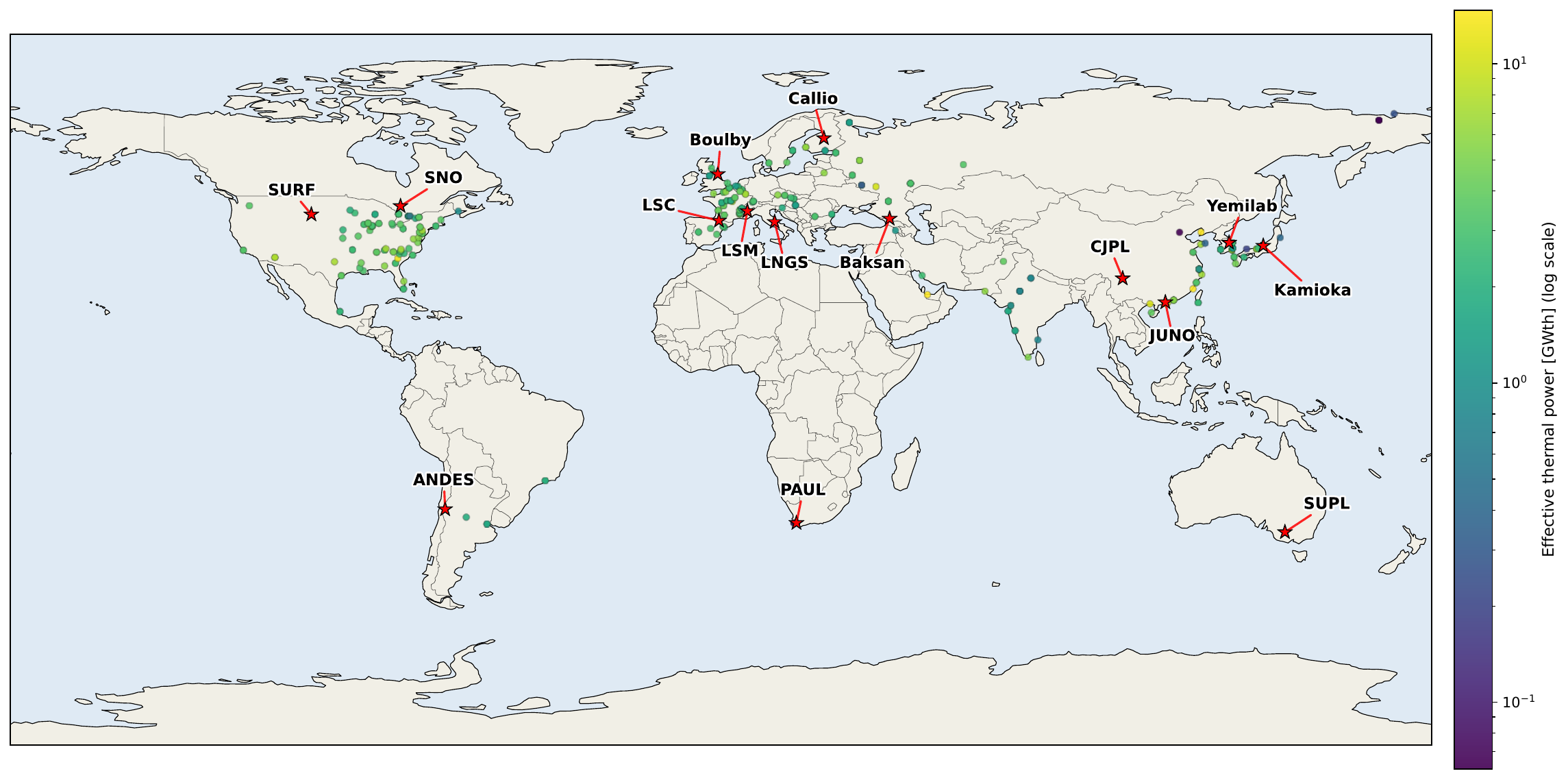}
    \caption{Worldwide distribution of operating reactor sites in 2024. The color bar indicates the annual-average effective thermal power of each reactor site, and the locations of underground laboratories are also shown. (The map was produced using Cartopy~\cite{cartopy}, with base features from Natural Earth.)}
    \label{fig:worldwide_reactors_labs}
\end{figure}

The Power Reactor Information System (PRIS)~\cite{PRIS}, established and maintained by the International Atomic Energy Agency (IAEA), is the most comprehensive and authoritative database of nuclear reactors worldwide. As of 2025, there are approximately 413 nuclear reactors in operation globally, with a total net installed capacity of 377.147 GW$_{\text{e}}$ (electrical power ); another 62 are under construction, with an installed capacity of 65.097 GW$_{\text{e}}$; more than 20 are in long-term shutdown (also referred to as suspended operation), with a total capacity of about 19.687 GW$_{\text{e}}$. Figure~\ref{fig:worldwide_reactors_labs} shows the worldwide nuclear reactor location distribution and the locations of underground labs. 

The numbers quoted above provide a general overview of the global reactor fleet as of 2025. For the flux calculations in this work, the baseline list of operating reactor units and their operational attributes is constructed from the 2024 operating fleet reported in the IAEA annual report \emph{Operating Experience with Nuclear Power Stations in Member States, 2025 Edition} (IAEA/OPEX/2025), which is a direct output of the IAEA Power Reactor Information System (PRIS)~\cite{IAEAOPEX2025, IAEAPRIS}. These unit-level records, including reactor types, electrical capacities, thermal powers, and load factors, are cross-checked using the INFN Ferrara reactor database published at the \emph{Antineutrino} website, which compiles PRIS-based worldwide reactor information on a yearly basis~\cite{INFNAntineutrinoDB}. Reactor-site coordinates are primarily taken from the INFN database and cross-checked against the WRI Global Power Plant Database. For the forward-looking reactor configurations used in the 2026 and 2030 predictions, information on construction progress and permanent shutdowns is compiled from the World Nuclear Association (WNA) Reactor Database and the Global Nuclear Power Tracker (GNPT) released by Global Energy Monitor~\cite{ WNAReactorDB, GEMGNPT}.

For a given prediction year (2026 and 2030), reactor units that are not operational, including long-term shutdown reactors and units that remain under construction, are assigned zero neutrino flux. For operational units, the production rate of $\bar{\nu}_e$'s is proportional to the reactor thermal power, $P_{\mathrm{th}}$ (GW$_{\mathrm{th}}$). Whenever a unit-level thermal-power value is available, we use the reported value directly. Otherwise, the thermal power is inferred from the electrical capacities, $P_{\mathrm e}$ (GW$_{\mathrm e}$), and reactor type using the reactor-type mean thermal efficiency listed in Table~\ref{table:conversion_efficiency},
\[
P_{\mathrm{th}}=\frac{P_{\mathrm e}}{\bar{\eta}_{\mathrm{type}}}.
\]
Table~\ref{table:conversion_efficiency} lists the thermal efficiencies of different reactor types, with typical values ranging from 0.31 to 0.43. Each GW$_{\text{th}}$ of thermal power corresponds to approximately $6.2\times10^{20}$ $\Bar{\nu}_e$'s per second, based on the average beta-decay spectrum of fission products in typical commercial reactors~\cite{Djurcic:2008ny, Vogel:2007du, Mueller:2011nm}. Additionally, given operational interruptions from refueling outages and maintenance, we introduce the reactor load factor (LF), defined as the ratio of the actual electrical energy generated over a given period to the reference energy that would be produced if the unit operated continuously at its reference power over the same period. For the baseline operating fleet (2024), 
the reactor neutrino flux is temporally averaged using the annual unit-level load factors reported in IAEA/OPEX/2025~\cite{IAEAOPEX2025}. For units newly entering the 2026 or 2030 prediction samples but lacking an observed LF in the baseline data, we adopt the reactor-type mean LF from Table~\ref{table:conversion_efficiency} as a proxy. 

\begin{table}[t]
    \centering
\begin{tabular}{l|c|c|c}\hline\hline
Reactor type  &  \# of units & $\bar{\eta}_{\mathrm{type}}$$\ \pm \ $SE & $LF$ (\%, mean $\pm$ SE)\\ \hline
PWR  & 306    & $0.352 \pm 0.001$   & $77.76 \pm 1.42$\\
PHWR & 45     & $0.317 \pm 0.003$   & $76.88 \pm 3.97$\\
BWR  & 43     & $0.342 \pm 0.002$   & $89.97 \pm 1.87$\\
LWGR & 10     & $0.312 \pm 0.002$   & $65.66 \pm 8.76$\\ 
GCR  & 8      & $0.428 \pm 0.003$   & $68.84 \pm 3.30$\\
FBR  & 2      & $0.415$   & $69.30$\\
HTGR & 1      & 0.42   & 20.66\\ \hline\hline
    \end{tabular}
 \caption{Average thermal efficiency $\eta$ and load factor $LF$ for various types of reactors in operation in 2024, the SE of the reactor-type mean is estimated via nonparametric bootstrap resampling of units within each reactor type (with replacement; $B=30000$, SE is not quoted for reactor types with too few operating units). There are seven types of reactors: Pressurized water reactor (PWR), Pressurized heavy water reactor (PHWR), Boiling water reactor (BWR),
 Light water cooled graphite moderated reactor (LWGR), Gas cooling reactor (GCR), 
 Fast breeder reactor (FBR), and High temperature gas cooling reactor (HTGR). Electrical and thermal capacities, load factors are compiled from IAEA/OPEX/2025~\cite{IAEAPRIS, IAEAOPEX2025}.}
    \label{table:conversion_efficiency}
\end{table}

%% file: sec3.tex
\section{\texorpdfstring{Reactor $\bar{\nu}_e$ spectral model}{Reactor anti-electron neutrino spectral model}}

\begin{figure}
    \centering
    \includegraphics[width=1.0\linewidth]{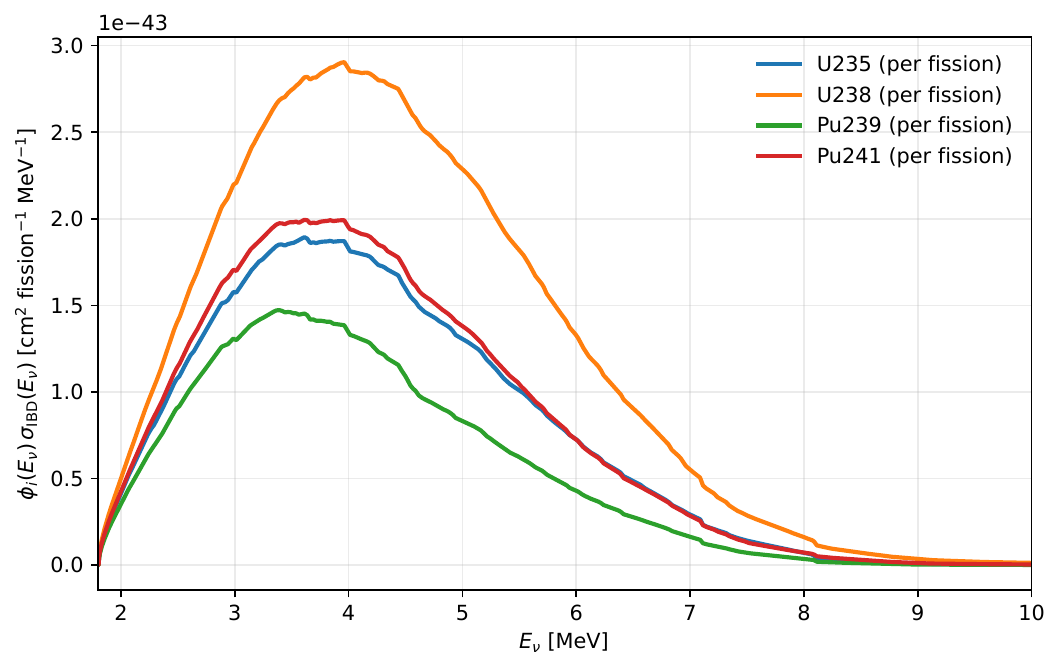}
    \caption{Detectable reactor $\bar{\nu}_e$ spectra of the four main fissile isotopes after weighting by the IBD cross section, computed with the SM2023 spectral model~\cite{SM2023}. The contributions from the four isotopes are shown separately.}
    \label{fig:reactor_neutrino_IBD}
\end{figure}

In the present work, the isotope-dependent reactor $\bar{\nu}_e$ spectra are modeled using SM2023~\cite{SM2023} for the four main fissile isotopes, $^{235}$U, $^{238}$U, $^{239}$Pu, and $^{241}$Pu. Since the spectral shapes differ among these isotopes, the adopted fission fractions affect the detectable spectrum after weighting by the IBD cross section. Figure~\ref{fig:reactor_neutrino_IBD} shows the corresponding isotope-resolved detectable spectra. The default reactor-side inputs adopted in the production calculations, including the fission fractions and their uncertainty treatment, are introduced later in Section~\ref{sec:default-setup}.

%% file: sec4.tex
\section{Pathlength and oscillation in vacuum}

\begin{figure}
    \centering
    \includegraphics[width=1.0\linewidth]{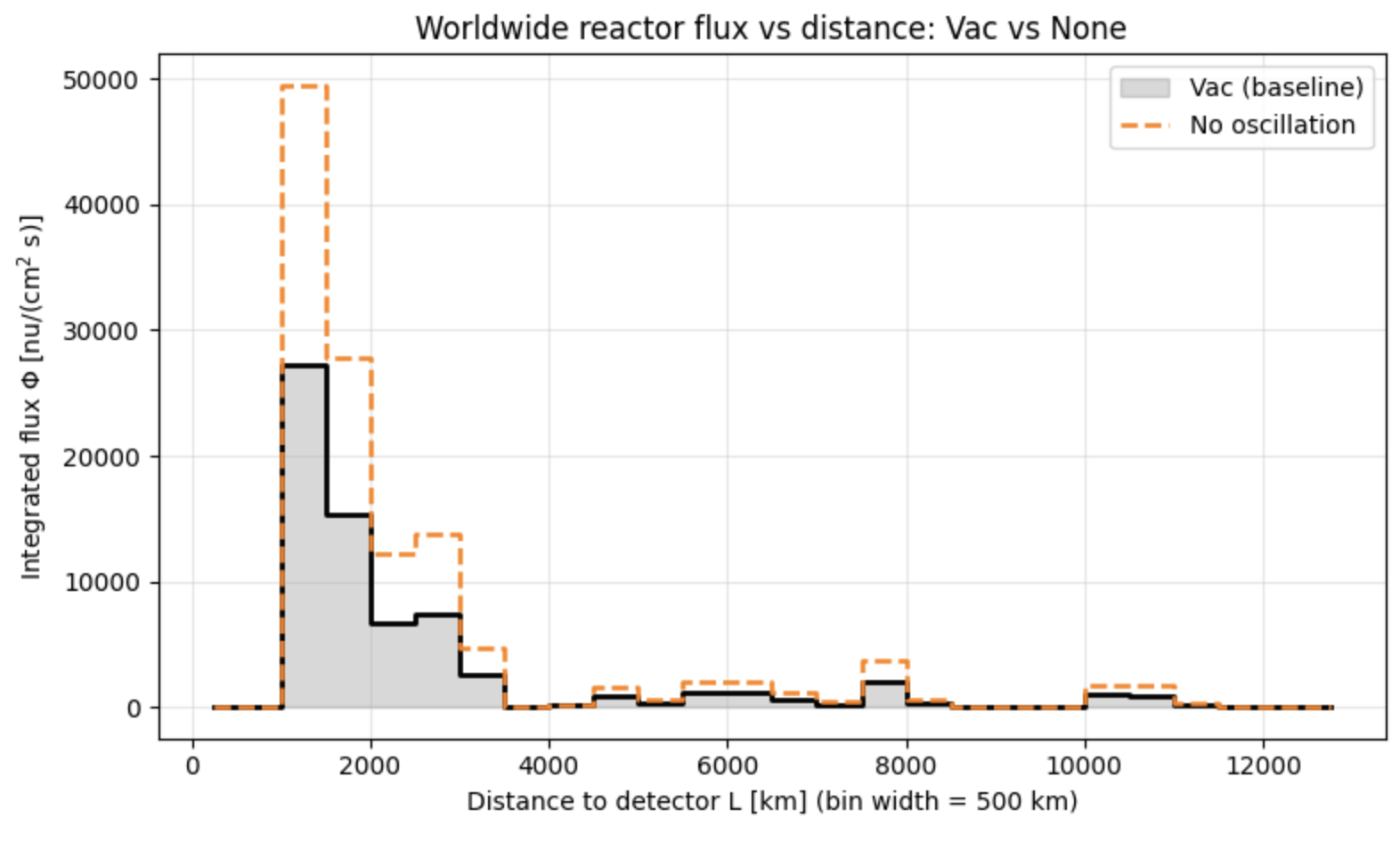}
    \caption{Reactor $\bar{\nu}_e$ flux varies with distance for CJPL.}
    \label{fig:flux_vs_distance}
\end{figure}

Using the reactor-site coordinates described in Section~\ref {sec:reactor_survey} and the coordinates for underground labs as given in 
Table~\ref{table:coordinates_labs}, we calculate the reactor-detector baseline assuming an Earth-centered spherical geometry. Neglecting the detector depth relative to the Earth's radius, we define the unit vectors pointing from the Earth's center to the detector and reactor sites as
\begin{align}
\mathbf{u}_{\mathrm{det}} &=
\left(
\cos\phi_{\mathrm{det}}\cos\lambda_{\mathrm{det}},\,
\cos\phi_{\mathrm{det}}\sin\lambda_{\mathrm{det}},\,
\sin\phi_{\mathrm{det}}
\right), \nonumber\\
\mathbf{u}_{\mathrm{reac}} &=
\left(
\cos\phi_{\mathrm{reac}}\cos\lambda_{\mathrm{reac}},\,
\cos\phi_{\mathrm{reac}}\sin\lambda_{\mathrm{reac}},\,
\sin\phi_{\mathrm{reac}}
\right), \nonumber\\
L &= R_{\oplus}\left\|\mathbf{u}_{\mathrm{det}}-\mathbf{u}_{\mathrm{reac}}\right\|,
\end{align}
where $R_{\oplus}=6371$ km is defined as the radius of the Earth, and $\phi_{\mathrm{det}}$, $\lambda_{\mathrm{det}}$, $\phi_{\mathrm{reac}}$, and $\lambda_{\mathrm{reac}}$ denote the latitudes and longitudes of the detector and reactor sites, respectively. Given the reactor-detector baseline $L$ and $\bar{\nu}_e$ energy $E_\nu$, the three-flavor $\bar{\nu}_e$ survival probability in vacuum is written as
\begin{align}
P_{ee}^{\mathrm{vac}}(E_\nu,L) =
1
&- \sin^2(2\theta_{12})\,c_{13}^4
\sin^2\!\left(\frac{\Delta m_{21}^2 L}{4E_\nu}\right) \nonumber\\
&- \sin^2(2\theta_{13})\,c_{12}^2
\sin^2\!\left(\frac{\Delta m_{31}^2 L}{4E_\nu}\right) \nonumber\\
&- \sin^2(2\theta_{13})\,s_{12}^2
\sin^2\!\left(\frac{\Delta m_{32}^2 L}{4E_\nu}\right),
\label{eq:vac_osc}
\end{align}
where $c_{ij}\equiv\cos\theta_{ij}$ and $s_{ij}\equiv\sin\theta_{ij}$. In the vacuum
case, the reactor $\bar{\nu}_e$ flux contribution from reactor $r$ to detector $d$ is
therefore proportional to $P_{ee}^{\mathrm{vac}}(E_\nu, L)/(4\pi L^2)$.
We take CJPL as an example for conducting the geoneutrino experiment, as it is sufficiently distant from commercial nuclear power plants at this time. 
Figure~\ref{fig:flux_vs_distance} shows the expected reactor $\bar{\nu}_e$ flux at CJPL as a function of source--detector distance, indicating that the dominant contributions arise from reactor complexes at baselines of roughly $10^3$ km, in particular Fangchenggang and Yangjiang power plants.

\begin{table}[t]
    \centering
\begin{tabular}{l|c|c|r}\hline\hline
Lab         & Latitude  & Longitude & Country\\ \hline
ANDES~\cite{civitarese2016andes}	    & $-30.25$	& $-69.883$   & Argentina \\ 
Baksan~\cite{karpov2006baksan_muonbursts}	    & 43.28	    & 42.69     & Russia \\
Boulby~\cite{vahsen2020cygnus}	    & 54.553	& $-0.825$    & UK\\
CallioLab~\cite{joutsenvaara2024callio}	& 63.659	& 26.042   & Finland\\
CJPL        & 28.2       & 101.7      & China \\
JUNO~\cite{an2016juno}        & 22.118     & 112.518    & China \\
Kamioka~\cite{futagami1999eastwest}	    & 36.426	& 137.310 & Japan\\
LNGS~\cite{bernabei2015dama_earthshadow}	    & 42.450	& 13.567  & Italy\\
LSC~\cite{trzaska2019canfranc_muon}         & 42.775	& $-0.529$  & Spain\\
LSM~\cite{ohare2015readout}         & 45.200   	& 6.670       & France\\
SNO~\cite{aharmim2013sno_phase3}	        & 46.475	    & $-81.201$ & Canada\\
SUPL~\cite{civitarese2016andes}	    & $-37.050$	& 142.767 & Australia\\
SURF~\cite{kennedy2015surf_microgravity}	    & 44.352	& $-103.752$  & USA\\
Yemilab~\cite{seo2021yemilab_darkphoton}	    & 37.189	& 128.659 & South Korea\\ \hline\hline
    \end{tabular}
 \caption{Coordinates for underground laboratories used in this work.
 There is no official report for the CJPL (China JinPing underground lab)
 coordinates, here the values are our best guess from the study of cosmic
 ray~\cite{JNE:2020bwn, JNE:2024gov}.
 }
    \label{table:coordinates_labs}
\end{table}

%% file: sec5.tex
\section{Neutrino propagation: vacuum vs. matter effect inside the Earth}
To describe the propagation of MeV-scale $\bar{\nu}_e$'s through the Earth, we first formulate the evolution equation in matter. This framework allows us to incorporate Earth-density profiles straightforwardly and to compare matter-modified propagation with the vacuum case.

\subsection{Evolution equation}
For a $\bar{\nu}_e$ propagating through the Earth, the MSW effect is described by the Schrödinger-like equation,
	\begin{equation}
 	i\,\frac{d}{dx}\,|\bar{\nu}_\text{base}(x)\rangle \;=\; H_\text{base}(x;E) \,|\bar{\nu}_\text{base}(x)\rangle,
 	\label{eq:schrodinger_evolution}
	\end{equation}
$H_\text{base}$ denotes the Hamiltonian matrix in a given basis ($\text{base} = \text{mass}:$ in vacuum mass basis; $\text{base} = \text{flavor}:$ in flavor basis), $E$ is the $\bar{\nu}_e$ energy, and $|\bar{\nu}_\text{base}(x)\rangle$ is the $\bar{\nu}_e$ state vector represented in that basis. For reactor $\bar{\nu}_e$'s, the initial condition at the source is $|\bar{\nu}_\text{mass}(x=0)\rangle = U^T(1,\ 0, \ 0)^T$, corresponding to a pure $\bar{\nu}_e$ in vacuum mass basis. It is convenient to evaluate the process by introducing an evolution operator defined as
	\begin{align}
 	&|\bar{\nu}_\text{base}(x)\rangle \;=\; S_\text{base}(x,x_0)\,|\bar{\nu}_\text{base}(x_0)\rangle,\quad S_\text{base}(x_0,x_0)=I,\nonumber\\
 	&i\,\frac{d}{dx}\,S_\text{base}(x,x_0) \;=\; H_\text{base}(x;E)\,S_\text{base}(x,x_0).
 	\label{eq:evolution_operator_def}
	\end{align}
$H_\text{base}(x; E)$ of different locations on the trajectory does not generally commute; therefore, the solution should be a path-ordered exponential written with the order of coordinates $\mathcal{P}$,
	\begin{equation}
 	S_\text{base}(L,0) \;=\; \mathcal{P}\exp\!\left[-i\int_{0}^{L} H_\text{base}(x;E)\,dx\right],
 	\label{eq:path_ordered_exp}
	\end{equation}
where $L$ is the total length of the trajectory. In practice, a valid approximation is to discretize the trajectory into many layer-steps and assume the matter potential is constant within each layer step. Under such an approximation, the evolution operator can be evaluated as,
	\begin{align}
	S_\text{base}(x_N=L,0) \;& \approx\; \prod_{k=N}^{1} U_k, \nonumber\\
    & = \prod_{k=N}^{1} \exp\!\left[-i\,H_{\text{base},k}\,\Delta x_k\right] \nonumber\\
	& = \exp(-iH_{N}\,\Delta x_{N})\cdots \exp(-iH_{2}\,\Delta x_{2})\exp(-iH_{1}\,\Delta x_{1}),
	\label{eq:layer_product}
	\end{align} 
 where $H_{\text{base},k}=H_\text{vac}(E)+H_{\text{mat},k}$, and $U_k$ is the evolution operator for the $k$th layer step. 
 Equation~\eqref{eq:layer_product} reduces the problem to the evaluation of a product of layer step evolution operators along the trajectory.
In the following, we present two computational approaches to compute this product after discretizing the trajectory into finite-layer steps.

\subsection{A numerical approach to the matter effect: the Strang-splitting method}
We first consider a numerical evaluation of Eq.~\eqref{eq:layer_product} in the vacuum mass basis, where $H_{\text{mass},k}=H_\text{vac}(E)+H_{\text{mat},k}$, with
\begin{align}
H_\text{vac}= \mathrm{diag}\left(0,\ \frac{\Delta m_{21}^2}{2E},\ \frac{\Delta m_{31}^2}{2E}\right),
\quad
H_{\text{mat},k}=U^T\mathrm{diag}(V^{CC}_k,\ 0,\ 0)U^*,
\end{align}
where $U$ is the PMNS matrix and $V^{CC}_k\equiv \eta \sqrt{2}G_F N_e(x_k)$ denotes the charged-current matter potential evaluated at the midpoint of layer step $k$, with $\eta=+1$ for neutrinos and $\eta=-1$ for $\bar{\nu}_e$'s. On this basis, the evolution operator for the $k$th layer step is
\begin{equation}
U_k=\exp\!\left[-i(H_{\rm vac}+H_{{\rm mat},k})\Delta x_k\right].
\end{equation}
To avoid repeated diagonalization of the full Hamiltonian $H_{\text{vac}}+H_{\text{mat},k}$ for each layer step, which becomes computationally expensive for a finely discretized trajectory, we apply a second-order Strang splitting to the layer step evolution operator,
\begin{equation}
\exp[-i(H_{\rm vac}+H_{{\rm mat},k})\Delta x_k]
\;\approx\;
\exp\left(\frac{-iH_{\rm vac}\Delta x_k}{2}\right)
\exp(-iH_{{\rm mat},k}\Delta x_k)
\exp\left(\frac{-iH_{\rm vac}\Delta x_k}{2}\right).
\label{eq:strang_layer}
\end{equation}
This decomposition achieves second-order accuracy globally with local error
 $\mathcal{O}(\Delta x_k^3)$ while preserving unitarity up to floating-point rounding.
Furthermore, $H_{\text{vac}}$ in the vacuum half-step term is already diagonal; the exponential form, therefore, can be written directly into the matrix, which implies 
\begin{equation}
\exp\left ({\frac{-iH_\text{vac} \Delta x_k}2}\right)=
\begin{pmatrix}
1&0 &0 \\
0 &\exp\left(\frac{-i\Delta m_{21}^2\Delta x_k}{4E}\right)  &0\\
0 & 0 &\exp\left(\frac{-i\Delta m_{31}^2\Delta x_k}{4E}\right)
\end{pmatrix},
\end{equation}
with a public phase subtracted. By defining $|e\rangle \equiv U^T|\bar{\nu}_e\rangle_\text{flavor}$ (in which $|\bar{\nu}_e\rangle_\text{flavor}=(1,\ 0,\ 0)^T$), which can be considered as the projection in the electron-flavor direction in the vacuum mass basis, the matter potential full-step term can be written in a closed form,
\begin{align}
\exp(-iH_{{\rm mat},k}\Delta x_k)
&= \exp[-iU^T \mathrm{diag}\!\left(V^{CC}_k,0,0\right)U^*\,\Delta x_k] \nonumber\\
&= \exp(-iV^{CC}_k\Delta x_k\,U^T |\bar{\nu}_e\rangle_f{\ }_f\langle \bar{\nu}_e|\,U^*)\nonumber\\
&= \exp(-iV^{CC}_k\Delta x_k\,|e\rangle\langle e|)\nonumber\\
&= I+[\exp(-iV^{CC}_k\Delta x_k)-1]P,
\label{eq:rank1_projector}
\end{align}
where we use the identity property of the projector $P \equiv |e \rangle \langle e|=P^n (\text{for }n \ge 1)$.
Therefore, the survival probability can be evaluated as,
\begin{align}
P_{ee} \simeq \left|{}_\text{mass}\langle \bar{\nu}_e|
\prod_{k=N}^1 e^{-iH_{\rm vac}\Delta x_k/2}\,
\left[I+\left(e^{-iV^{CC}_k\Delta x_k}-1\right)P\right]\,
e^{-iH_{\rm vac}\Delta x_k/2}
|\bar{\nu}_e\rangle_\text{mass} \right|^2
\label{eq:strang_P}
\end{align}
At each layer step, the computation involves only the scalar quantity $V_k^{CC}\Delta x_k$ and the fixed projector $P$; the vacuum phase factors are the same for all trajectories and can be precomputed before the propagation is solved. To obtain a further optimalization, Eq.~\eqref{eq:strang_P} can be rewritten by combining the neighbor vacuum half steps between layer steps $e^{-iH_{\rm vac}\Delta (x_{k+1}+x_{k})/2}$, which can slightly reduce the computational load for the matrices production of each layer step.
This makes the algorithm computationally inexpensive and easy to parallelize over energies or event samples. As a complementary treatment to this numerical scheme, we next consider a first-order analytical approximation to the MSW effect.

\subsection{First-order analytical approximation to the matter effect}
As a complementary approach to the numerical Strang-splitting method above, we also derive a first-order approximation to the MSW effect in Appendix~\ref{sec:1st_order}, following the expansion in the layer-to-layer rotation of the $1$--$2$ matter mixing angle in Ref.~\cite{Ioannisian:2017dkx}. We obtain both an effective $SU(2)$ expansion and a full $SU(3)$ expansion. In the main text, we quote only the $SU(3)$ result, since it retains the complete three-flavor structure, while the $SU(2)$ expression is recovered after averaging over the unresolved $1$--$3$ oscillations. The resulting three-flavor survival probability can be written as,
\begin{equation}
P^\mathrm{3fl}_{ee}\simeq|A^{\mathrm{3fl},0}_{ee}|^2+2\,\Re[(A^{\mathrm{3fl},0}_{ee})^* \cdot A^{\mathrm{3fl},1}_{ee}],
\label{eq:S_probibility}
\end{equation}
where 
\begin{align}
|A^{\mathrm{3fl},0}_{ee}|^2 =1 &-\sin^2(2\theta_{12})c_{13}^4\sin^2(\frac{\phi^\text{tot}_{21}}{2}) \nonumber\\
&-\sin^2(2\theta_{13})c_{12}^2\sin^2(\frac{\Delta m^2_{31}L}{4E}) \nonumber\\
&-\sin^2(2\theta_{13})s_{12}^2\sin^2(\frac{\Delta m^2_{31}L}{4E}-\frac{\phi^\text{tot}_{21}}{2}),
\label{eq:S_probibility_1}
\end{align}
and
\begin{align}
2\,\Re[&(A^{\mathrm{3fl},0}_{ee})^*\cdot A^{\mathrm{3fl},1}_{ee}]
= -\sin^2(2\theta_{12})c_{13}^{2} \nonumber\\
&\quad \times
\Bigg[
c_{13}^{2}\cos(2\theta_{12})\,\sin\!\left(\frac{\phi^\text{tot}_{21}}{2}\right)
+ s_{13}^{2}\,\sin\!\left(\frac{\phi^\text{tot}_{21}}{2}-\frac{\Delta m^{2}_{31}L}{2E}\right)
\Bigg] \nonumber\\
&\quad \times \int_{0}^{L}V_{\mathrm{eff}}(x)\,
\cos\!\left(\frac{\phi^\text{tot}_{21}}{2}-\phi^m_{x \rightarrow L}(x)\right)dx.
\label{eq:S_probibility_2}
\end{align}
with $c_{12}\equiv \cos(\theta_{12})$ and $s_{12}\equiv \sin(\theta_{12})$, the term $V_{\mathrm{eff}}=c^2_{13}V^{CC}$, is the matter potential projected onto the effective $(1$--$2)$ subsystem. By setting $V_{\rm eff}(x)\to0$, the term $\phi^\text{tot}_{21}$ (Eq.~\ref{eq:phi_tot}) reduces to $\Delta m^2_{21}L/2E$, therefore the whole equation reduces to the standard vacuum oscillation Eq.~\eqref{eq:vac_osc}.

In Section~\ref{sec:converge_test}, we compare the convergence of the first-order approximation and the Strang-splitting method as the layer step size is reduced. In the numerical calculations presented there, each trajectory is discretized into $8000$ uniform layer steps for the MSW evolution.

%% file: sec6.tex


\section{Density models for the Earth}
Accounting for Earth's density profile in neutrino oscillations requires a refined Earth-density model. 
In density models, we consider one-dimensional (1D) and three-dimensional (3D) models for analytical calculations, which are robust enough to accommodate more refined Earth models.  

\subsection{The 1D radial reference model}
\label{sec:prem_1d}

As a 1D spherically symmetric baseline for Earth matter effects, we adopt the Preliminary Reference Earth Model (PREM) by Dziewonski and Anderson \cite{PREM}, which provides the radially varying mass density profile $\rho(r)$. In our implementation, the tabulated PREM density is converted to depth $d = R_\oplus - r$ and evaluated along neutrino trajectories via linear interpolation; values outside the tabulated range are clamped to the nearest endpoint density to avoid extrapolation artifacts.

To compute the charged-current matter potential, we use a simplified piecewise-constant electron fraction $Y_e(d)$:
\begin{equation}
Y_e(d)=
\begin{cases}
0.50, & d < 2891~\mathrm{km} \quad (\text{mantle}),\\
0.467, & d \ge 2891~\mathrm{km} \quad (\text{core}),
\end{cases}
\end{equation}
and evaluate $V_e(d)$ from $Y_e(d)$ and the PREM density profile (with $V_e \rightarrow -V_e$ for $\bar{\nu}_e$'s).

\subsection{The 3D Earth density model construction}
To analyze the density heterogeneity of the Earth, we construct a hybrid 3D model, dividing the Earth into the shallow layer ($0-320$ km, from the crust to the upper-most mantle), the mantle layer ($320-2880$ km), and the Deep Earth layer ($2880-6371$ km, mainly the Earth's core). For each layer, we adopt a different model to describe the density trait.
\begin{itemize}
\item {\textbf{LITHO1.0 (0--320 km):}} 
LITHO 1.0 is a $1^\circ \times 1^\circ$-resolution model of the crust and uppermost mantle with a layered parameterization. Layer-wise physical properties (including density and material type) are provided explicitly, such as water/ice, sediments, crystalline crust, and the lithospheric mantle ~\cite{LITHO10}. To construct the matter potential model, we use the LITHO1.0 density field in this depth range and construct a corresponding electron-fraction field by assigning a constant $Y_e$ to each material layer:
\begin{equation}
Y_e =
\begin{cases}
0.555, & \text{water or ice},\\
0.50,  & \text{sediments, crust, and lithospheric plates
}
\end{cases}
\end{equation}
The local matter potential is evaluated from the product $Y_e \rho$ (with $V_e \rightarrow -V_e$ for $\bar{\nu}_e$'s).

\item {\textbf{SAW642AN (320--2880 km):}}
SAW642AN is a global whole-mantle radially anisotropic shear-velocity model. In terms of formulation, only the isotropic shear velocity $V_S$ (Voigt average) and the anisotropic parameter $\xi$ are inverted, while other quantities such as density are tied to $V_S$ through empirical scaling relations ~\cite{SAW642AN},
\begin{equation}
\frac{\delta \ln \rho}{\delta \ln V_S} = 0.33 \, ,
\end{equation}
in practice we use the density field $\rho$ provided in the distributed SAW642AN (IRIS) product.
For the electron fraction in the mantle, we adopt a constant value $Y_e = 0.50$.

\item {\textbf{Deep Earth layer (beyond 2880 km):}}
The 3D mantle models used in this work do not extend into the Earth's core, since no mature model describes the density of the deep-Earth layer beyond 2880 km (primarily the core). Therefore, for depths below the deepest 3D coverage, we continue to use the 1D PREM radial profile for density, as described in Section~\ref{sec:prem_1d} \cite{PREM}, and adopt the same piecewise electron-fraction prescription (mantle $Y_e=0.50$, core $Y_e=0.467$).
\end{itemize}

%% file: sec7.tex
\section{Flux and spectrum predictions: 1D vs. 3D models}

In this section, we will incorporate geophysical models into the analysis of reactor $\bar{\nu}_e$ propagation through Earth's interior to assess the influence of matter effects on the predicted $\bar{\nu}_e$ flux and energy spectrum at underground labs. We will also present our results under the vacuum-oscillation hypothesis to validate the calculation.  

\subsection{Default parameters setup and uncertainty propagation}
\label{sec:default-setup}
In this work, the Summation Model (SM2023)~\cite{SM2023} is used to model the $\bar{\nu}_e$ energy spectrum. The default fission fractions $f_i$ adopted for different reactor classes are listed in Table~\ref{table:fission_factor}. 
For pressurized heavy-water reactors (PHWRs), we adopt the fission fractions from the SNO+ measurement~\cite{SNO:2025PRL}. 
For all other reactor classes, we follow the reference values in Baldoncini \textit{et al.}~\cite{Baldoncini2015}. The mean energies released per fission~\cite{Baldoncini2015} are
$\langle Q_i\rangle = \{\,{}^{235}\mathrm{U}:202.36~{\pm ~0.26},\ {}^{238}\mathrm{U}:205.99~{\pm ~0.52},\ {}^{239}\mathrm{Pu}:211.12~{\pm ~0.34},\ {}^{241}\mathrm{Pu}:214.26~{\pm ~0.33}\,\}\,\mathrm{MeV/fission}$.
The oscillation parameters are taken from the PDG 2025 review of neutrino masses, mixing, and oscillations, adopting the normal-ordering global-fit values summarized in Table~14.7~\cite{PDG2024, PDG2025NuMix}: 
$\sin^2\theta_{12}=0.308^{+0.012}_{-0.011}$, 
$\sin^2\theta_{13}=2.215^{+0.056}_{-0.058}\times 10^{-2}$, 
$\Delta m^2_{21}=(7.49^{+0.19}_{-0.20})\times10^{-5}\,\mathrm{eV}^2$, 
and 
$\Delta m^2_{32}=(2.438^{+0.021}_{-0.019})\times10^{-3}\,\mathrm{eV}^2$. 
The corresponding $3\sigma$ ranges are 
$0.275$--$0.345$, 
$0.02030$--$0.02388$, 
$(6.92$--$8.05)\times10^{-5}\,\mathrm{eV}^2$, 
and 
$(2.376$--$2.503)\times10^{-3}\,\mathrm{eV}^2$, respectively.
The IBD total cross section is evaluated using the analytical approximation of Strumia and Vissani~\cite{sv_2003}.

To evaluate the uncertainty propagation in this work, we follow the general Monte Carlo strategy of Baldoncini \textit{et al.}~\cite{Baldoncini2015}, namely to propagate the relevant nuisance inputs through repeated evaluations of the reactor-$\bar{\nu}_e$ prediction. In comparison with Baldoncini \textit{et al.}, the present implementation is extended to include the spectral uncertainty of SM2023 and the updated oscillation-input uncertainties.
A direct joint Monte Carlo propagation of all reactor-side and oscillation-side nuisance parameters in the full three-flavor MSW calculation would require an impractically high computational cost. 
Schematically, the predicted observable can be written as
\begin{equation}
X(\boldsymbol{\xi}_{\rm reac},\boldsymbol{\xi}_{\rm osc})
=
\sum_r \int dE_\nu\,
\mathcal{W}_r(E_\nu;\boldsymbol{\xi}_{\rm reac})\,
P_{ee}(E_\nu,\Gamma_r;\boldsymbol{\xi}_{\rm osc}),
\label{eq:unc_factorized_generic}
\end{equation}
where $X$ denotes either the integrated flux or the TNU prediction, $r$ runs over reactor units, $\Gamma_r$ denotes the matter profile along the $r$th trajectory, and $\boldsymbol{\xi}_{\rm reac}$ and $\boldsymbol{\xi}_{\rm osc}$ represent the reactor-side and oscillation-side nuisance parameters, respectively. Therefore, the uncertainty propagation is factorized into an oscillation sector and a reactor sector, which are evaluated separately and then combined by a final nonparametric resampling of the two response distributions. In particular, for the oscillation sector, the oscillation parameters are sampled by split-normal distributions truncated at the quoted $3\sigma$ ranges. There are $1500$ MC samples taken for the MSW calculation and $10000$ samples for the vacuum oscillation. 

Regarding the reactor sector, the SM2023 spectral uncertainty is propagated by adopting four-isotope spectrum from the covariance matrix~\cite{SM2023}; all reactors then share the sampled spectrum; the combined term $LF\times P_{\rm th}$ is assigned by a relative uncertainty of $2\%$, while an additional uncertainty of $0.4\%$ is included for the IBD cross-section~\cite{Baldoncini2015}. The mean energies released per fission are varied according to the uncertainties quoted above. Compared with Baldoncini \textit{et al.}~\cite{Baldoncini2015}, the treatment of the fission-fraction uncertainty is modified in the present work. Instead of discrete sampling from the literature fraction sets, we adopt the class-dependent default fractions listed in Table~\ref{table:fission_factor} as the nominal values, estimate the corresponding effective widths from the spread of the Baldoncini compilation, and propagate the uncertainty through continuous Gaussian perturbations with the normalization constraint $\sum_i f_i = 1$ imposed in each realization. The reactor sector uncertainty is propagated with $3\times10^5$ Monte Carlo realizations.

After a final nonparametric resampling procedure, the quoted asymmetric uncertainties correspond to the differences between the nominal central value and the 16th and 84th percentiles of the combined distribution, i.e., an approximate $68\%$ central interval. It should be emphasized that the quoted asymmetric uncertainty bands are obtained only from the Monte Carlo propagation of the oscillation-side and reactor-side input uncertainties. The shifts associated with the MSW solver, the adopted Earth model, and the neutrino mass ordering are not folded into the quoted bands. Instead, we report them separately through
diagnostic systematic shifts arising from the numerical treatment of the MSW effect within the 3D Earth model are quantified as
$\Delta_{\rm \mathcal{M}}\equiv 2(\mathrm{TNU}_{\rm Strang}-\mathrm{TNU}_{\rm 1st})/(\mathrm{TNU}_{\rm Strang}+\mathrm{TNU}_{\rm 1st})$,
where $\mathrm{TNU}_\mathrm{Strang}$ and $\mathrm{TNU}_\mathrm{1st}$ denote the total normalized unit predictions obtained using the Strang-splitting and first-order methods, respectively. Systematic uncertainty associated with Earth model assumptions, the difference between 3D and 1D Earth density profiles is defined as 
$\Delta_{\rm \oplus}\equiv 2(\mathrm{TNU}_{\rm 3D}-\mathrm{TNU}_{\rm 1D})/(\mathrm{TNU}_{\rm 3D}+\mathrm{TNU}_{\rm 1D})$.
Finally, systematic uncertainty due to the choice of neutrino mass ordering—normal (NO) versus inverted (IO) is expressed as $\Delta_{\rm N/I}\equiv 2(\mathrm{TNU}_{\rm NO}-\mathrm{TNU}_{\rm IO})/(\mathrm{TNU}_{\rm NO}+\mathrm{TNU}_{\rm IO})$
, with all calculations performed using the Strang-splitting method and the 3D Earth model.

In particular, since no robust probabilistic prescription is assigned to the terrestrial matter-profile uncertainty in the present work, $\Delta_{\oplus}$ should be interpreted as an indicator of Earth-model dependence rather than as a rigorously quantified statistical uncertainty.
\begin{table}[!htbp]
  \centering
  \caption{Default fission fractions $f_i$ adopted for different reactor classes.
  The isotopes are ordered as $^{235}\mathrm{U}$, $^{238}\mathrm{U}$, $^{239}\mathrm{Pu}$, and $^{241}\mathrm{Pu}$, and each row is normalized to unity.}
  \begin{tabular}{lcccc}
    \hline\hline
    Reactor class & $f_{235}$ & $f_{238}$ & $f_{239}$ & $f_{241}$ \\
    \hline
    PWR  & 0.606 & 0.074 & 0.274 & 0.046 \\
    BWR  & 0.488 & 0.087 & 0.359 & 0.067 \\
    LWGR & 0.580 & 0.074 & 0.292 & 0.054 \\
    GCR  & 0.544 & 0.075 & 0.318 & 0.063 \\
    PHWR & 0.520 & 0.050 & 0.420 & 0.010 \\
    MOX  & 0.000 & 0.083 & 0.708 & 0.209 \\
    \hline\hline
  \end{tabular}
  \label{table:fission_factor}
\end{table}

\subsection{The flux and IBD event rate predictions}
Due to the 2400 m rock overburden, CJPL has an advantage in carrying on a MeV-scale neutrino experiment. 
Also, it is far from commercial nuclear power plants and the nearby Qinghai-Tibet Plateau, making it an ideal location for a geoneutrino experiment. Below, 
we present the predictions for CJPL and other underground labs separately. 

\subsubsection{Predictions for CJPL}

\begin{figure}[t]
    \centering
    \begin{minipage}[t]{0.49\linewidth}
        \centering
        \includegraphics[width=\linewidth]{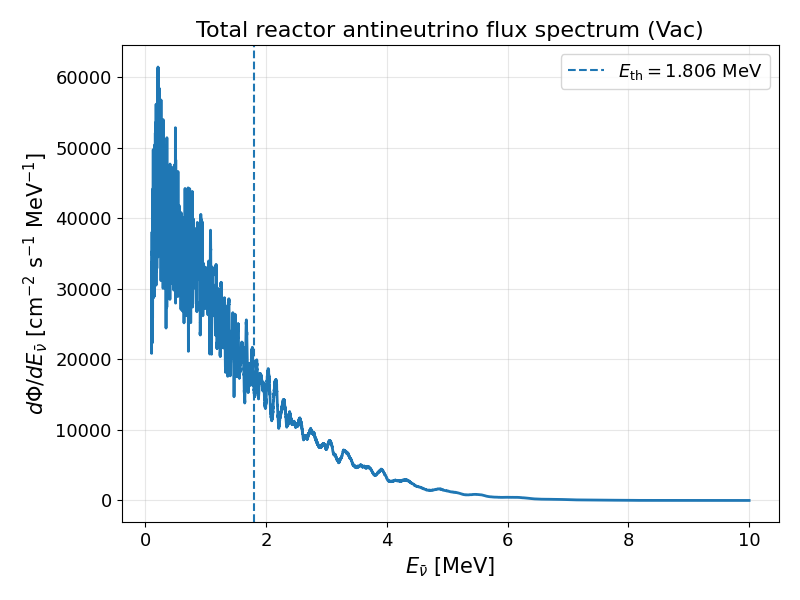}
    \end{minipage}\hfill
    \begin{minipage}[t]{0.49\linewidth}
        \centering
        \includegraphics[width=\linewidth]{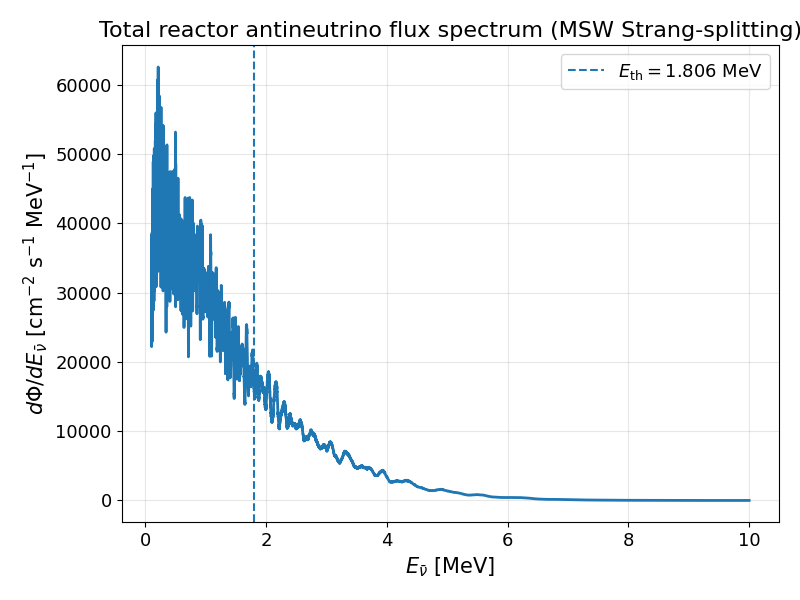}
    \end{minipage}
    \caption{Predicted reactor $\bar{\nu}_e$ flux spectrum at CJPL from the adopted global reactor sample. The left panel uses the vacuum survival probability in Eq.~\eqref{eq:vac_osc}. In contrast, the right panel includes the three-flavor MSW matter effect in the Earth evaluated with the Strang-splitting solver Eq.~\eqref{eq:strang_P}.}
    \label{fig:Jinping_flux_compare}
\end{figure}

Figure~\ref{fig:Jinping_flux_compare} compares the predicted reactor $\bar{\nu}_e$ flux spectrum at CJPL under the vacuum and MSW oscillation treatments. The overall spectral shapes are very similar, and the MSW correction introduces only a small energy-dependent modification. Using the uncertainty propagation described in Section~\ref{sec:default-setup}, the total flux integrated over $0.1$--$10~\mathrm{MeV}$ increases from
\[
\Phi_{\rm vac}=7.46^{+0.19}_{-0.21}\times10^{4}~\mathrm{cm}^{-2}\,\mathrm{s}^{-1}
\]
to
\[
\Phi_{\rm MSW}=7.48^{+0.19}_{-0.21}\times10^{4}~\mathrm{cm}^{-2}\,\mathrm{s}^{-1},
\]
corresponding to an increase of about $0.3\%$. This indicates that the matter effect yields a subpercent correction to the CJPL flux prediction in the reactor $\bar{\nu}_e$ energy range considered here.

\begin{figure}[t]
    \centering
    \begin{minipage}[t]{0.49\linewidth}
        \centering
        \includegraphics[width=\linewidth]{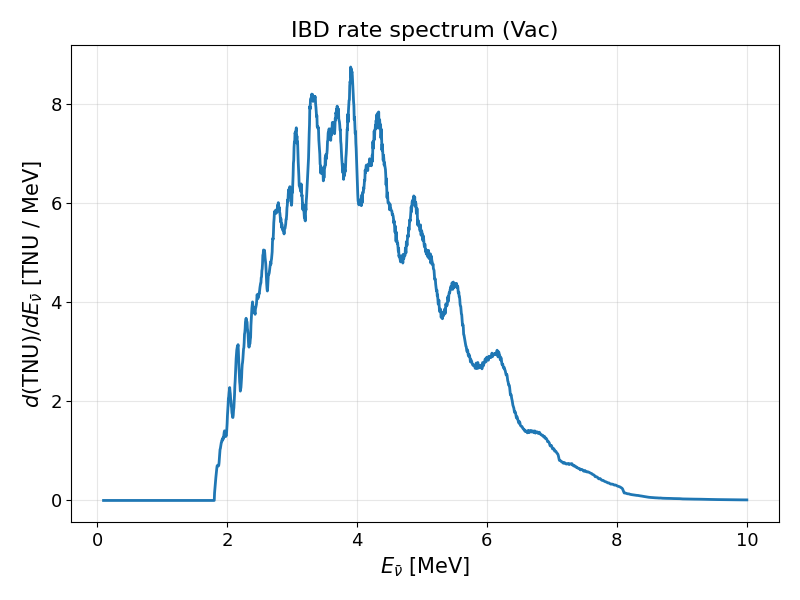}
    \end{minipage}\hfill
    \begin{minipage}[t]{0.49\linewidth}
        \centering
        \includegraphics[width=\linewidth]{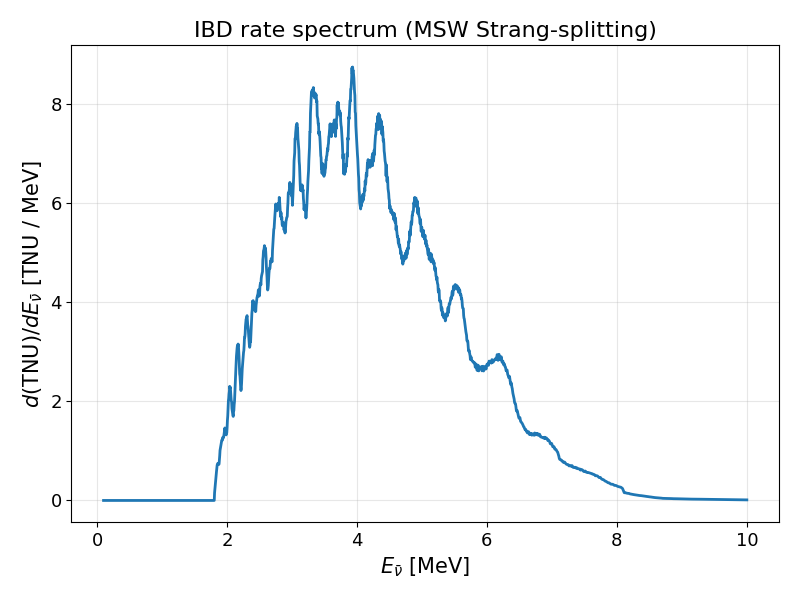}
    \end{minipage}
    \caption{Predicted IBD event rate spectrum at CJPL from the adopted global reactor sample, expressed in TNU. The left panel uses the vacuum survival probability Eq.~\eqref{eq:vac_osc}, while the right panel includes the three-flavor MSW matter effect in the Earth evaluated with the Strang-splitting solver Eq.~\eqref{eq:strang_P}.}
    \label{fig:Jinping_TNU_compare}
\end{figure}

Figure~\ref{fig:Jinping_TNU_compare} compares the predicted IBD event rate spectrum at CJPL under the vacuum and MSW oscillation treatments. The overall spectral shapes remain very similar, while the MSW correction produces a small energy-dependent enhancement above threshold. Using the uncertainty propagation described in Section~\ref{sec:default-setup}, the corresponding IBD event rate for $E_\nu>1.806~\mathrm{MeV}$ increases from
\[
R_{\rm IBD}^{\rm vac}=24.14^{+0.84}_{-0.83}\ \mathrm{TNU}
\]
to
\[
R_{\rm IBD}^{\rm MSW}=24.30^{+0.84}_{-0.84}\ \mathrm{TNU},
\]
corresponding to an increase of about $0.7\%$. Therefore, although the matter effect is still smaller than the current propagated uncertainty, it already reaches the sub-percent level and should be included in future precision reactor-background predictions for CJPL.

\subsubsection{Predictions for other underground labs}
In this section, we present predictions for the $\bar{\nu}_e$ flux and IBD event rate in Table~\ref{table:prediction_2026} and Table~\ref{table:prediction_2030}, at multiple underground labs mentioned in Table~\ref{table:coordinates_labs} for the years 2026 and 2030. The absolute reactor background differs strongly among underground labs, primarily reflecting the geographical distribution of the surrounding reactor fleet. JUNO and Yemilab receive more reactor neutrinos, while Boulby is also among the most in Europe. By contrast, SUPL, ANDES, and CJPL remain comparatively clean in absolute flux and TNU. 

From 2026 to 2030, most sites exhibit only moderate changes, but several Asian labs, especially CJPL and Kamioka, show a more visible increase under the adopted forward-year reactor configuration; SURF provides a mild exception to this trend, showing a slight decrease from 2026 to 2030. Within the adopted 2030 reactor-status configuration, this behavior is primarily attributable to the reduced contribution from the Diablo Canyon units, whose currently authorized operation extends only to late 2029 (Unit~1) and late 2030 (Unit~2). 

\begin{figure}[t]
    \centering
    \includegraphics[width=0.7\linewidth]{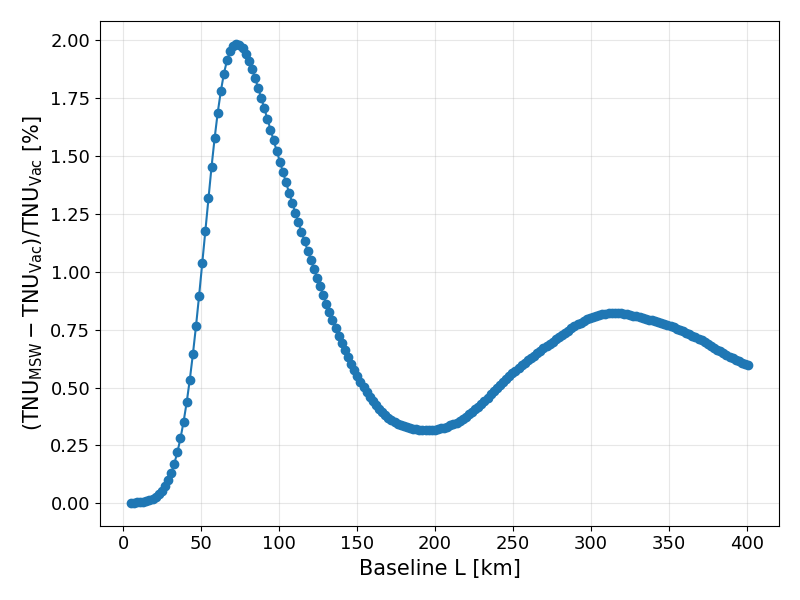}
    \caption{Relative MSW-induced enhancement of the single-baseline TNU contribution, $[{\rm TNU}_{\rm MSW}(L)-{\rm TNU}_{\rm vac}(L)]/{\rm TNU}_{\rm vac}(L)$, as a function of reactor--detector baseline $L$. Here, the Fangchenggang-1 reactor unit is used as an illustrative example.}
    \label{fig:msw-vac_rel}
\end{figure}

For the comparison with the prediction of vacuum and MSW effects, the flux differences are generally at the $10^{-4}$--$10^{-3}$ level, while the corresponding shifts in TNU are at the $10^{-3}$--$10^{-2}$ level. The solver dependence $\Delta_{\rm \mathcal{M}}$ and the mass-ordering dependence $\Delta_{\rm N/I}$ remains much smaller, typically at or below the few-$10^{-4}$ level. The Earth-model dependence $\Delta_{\oplus}$ is clearly the dominant diagnostic systematic among the three, ranging from about $0.05\%$ to $0.53\%$. Among all sites, JUNO and Yemilab exhibit a large vacuum-to-MSW shift in TNU, and also the largest Earth-model dependence $\Delta_{\oplus}$. In contrast, Boulby gives the smallest vacuum-to-MSW shifts in TNU and also the smallest Earth-model dependence. The Earth-model dependence is strongly correlated with the overall size of the MSW correction, but it is not determined solely by the contribution concentration. 
This behavior is consistent with the baseline dependence of the relative MSW-induced TNU enhancement shown in Fig.~\ref{fig:msw-vac_rel}. The enhancement is already appreciable in the medium-baseline region relevant for JUNO and Yemilab, but remains much smaller at the shorter baseline characteristic of the dominant Boulby contribution.
JUNO and Yemilab exhibit large $\Delta_{\oplus}$ because their reactor backgrounds are dominated by a few medium-baseline groups, such as the $\sim 52$ km (JUNO) and $\sim 66$ km (Yemilab) groups, for which the MSW correction is already appreciable and therefore also more sensitive to the adopted Earth model. By contrast, Boulby is also strongly dominated by a single baseline group. Still, this lab lies at a much shorter baseline ($\sim 24$ km), for which the accumulated matter effect remains intrinsically small. As a result, its dependence on the Earth model is much weaker despite the large reactor neutrino 
flux. For other labs such as LSM, SNO, and CJPL, the reactor neutrinos are distributed over a broader set of baseline groups, so the path-dependent matter corrections are accumulated less coherently in the total prediction. Their vacuum-to-MSW shifts therefore remain at an intermediate level, rather than reaching the larger values observed at JUNO and Yemilab.

\begin{table}[!htpb]
\centering
\small
\setlength{\tabcolsep}{4pt}     
\renewcommand{\arraystretch}{1.1}

\caption{Predicted integrated reactor $\bar{\nu}_e$ flux $\Phi$ and IBD event rate for underground laboratories in 2026. The MSW columns adopt the three-flavor matter-enhanced neutrino-oscillation formalism, incorporating a three-dimensional Earth-density model and employing the Strang-splitting method as the baseline computational approach. 
$\Phi\,[\mathrm{cm^{-2}\,s^{-1}}]$ is integrated over $E_\nu\in[0.1,10]$ MeV. The definitions for
systematic errors $\Delta_{\rm \mathcal{M}}$, $\Delta_{\rm \oplus}$, and $\Delta_{\rm N/I}$ are given in the text.
}

\begin{tabular}{
l
c
c
c
c
c
c
c
}
\toprule
& \multicolumn{2}{c}{Vacuum} & \multicolumn{2}{c}{MSW} & \multicolumn{3}{c}{|Syst.|[\%]} \\
\cmidrule(lr){2-3}\cmidrule(lr){4-5}\cmidrule(lr){6-8}
Site &
{$\Phi$ [$\times10^{4}$]} &
{TNU} &
{$\Phi$ [$\times10^{4}$]} &
{TNU} &
{$\Delta_{\rm \mathcal{M}}$ } &
{$\Delta_{\rm \oplus}$ } &
{$\Delta_{\rm N/I}$ }\\
\midrule
ANDES        &$1.46^{+0.04}_{-0.04}$&$4.64^{+0.16}_{-0.16}$&$1.46^{+0.04}_{-0.04}$&$4.67^{+0.12}_{-0.12}$&$0.01$&$0.19$&$0.02$   \\
Baksan       &$13.64^{+0.33}_{-0.41}$&$43.97^{+1.55}_{-1.53}$&$13.66^{+0.34}_{-0.40}$&$44.23^{+1.57}_{-1.51}$
&$0.00$
&$0.32$&$0.02$ \\
Boulby       &$248.01^{+8.20}_{-8.11}$&$1074.93^{+40.65}_{-40.02}$&$248.36^{+8.29}_{-7.96}$&$1075.99^{+40.24}_{-40.50}$ 
&$0.00$
&$0.05$&$0.02$       \\
CallioLab    &$22.20^{+0.66}_{-0.54}$&$72.10^{+2.50}_{-2.51}$&$22.25^{+0.66}_{-0.53}$&$72.55^{+2.54}_{-2.49}$&
$0.00$
&$0.35$&$0.01$\\
CJPL         &$7.46^{+0.19}_{-0.21}$&$24.14^{+0.84}_{-0.83}$&$7.48^{+0.19}_{-0.21}$ &$24.30^{+0.84}_{-0.84}$&$0.02$&$0.22$&$0.03$    \\
JUNO         &$669.58^{+19.75}_{-20.91}$&$1367.09^{+75.55}_{-69.99}$&$671.51^{+19.84}_{-20.79}$&$1379.98^{+78.21}_{-71.11}$&$0.02$&$0.53$&$0.03$  \\
Kamioka      &$58.37^{+1.62}_{-1.56}$&$199.13^{+6.86}_{-6.84}$&$58.48^{+1.63}_{-1.54}$&$199.90^{+6.90}_{-6.82}$&$0.01$&$0.22$&$0.02$   \\
LNGS         &$21.46^{+0.53}_{-0.61}$&$69.37^{+2.40}_{-2.39}$&$21.48^{+0.55}_{-0.59}$&$69.77^{+2.43}_{-2.38}$&$0.01$&$0.34$& $0.02$  \\
LSC &$66.81^{+1.62}_{-1.98}$&$218.71^{+7.57}_{-7.44}$&$66.94^{+1.65}_{-1.95}$&$219.75^{+7.63}_{-7.42}$&$0.01$&$0.31$&$0.02$ \\
LSM   &$137.49^{+3.45}_{-3.90}$&$474.23^{+16.09}_{-16.51}$&$137.79^{+3.48}_{-3.85}$&$476.88^{+16.30}_{-16.44}$&$0.01$&$0.35$&$0.02$ \\
SNO          &$55.93^{+1.61}_{-1.37}$&$175.34^{+6.21}_{-6.07}$&$56.04^{+1.63}_{-1.35}$&$176.37^{+6.30}_{-6.03}$&$0.01$&$0.36$&$0.02$ \\
SUPL         &$0.78^{+0.03}_{-0.01}$&$2.53^{+0.09}_{-0.09}$&$0.79^{+0.02}_{-0.02}$&$2.55^{+0.09}_{-0.09}$&$0.01$&$0.02$&$0.00$
\\
SURF         &$9.21^{+0.25}_{-0.24}$&$29.69^{+1.02}_{-1.02}$&$9.24^{+0.25}_{-0.24}$&$29.90^{+1.04}_{-1.02}$&$0.03$&$0.25$&$0.03$ \\
Yemilab      &$432.53^{+11.10}_{-13.13}$&$908.08^{+43.71}_{-41.78}$&$433.45^{+11.34}_{-12.85}$&$917.77^{+45.20}_{-40.68}$&$0.01$&$0.51$&$0.03$  \\
\bottomrule
\end{tabular}

\label{table:prediction_2026}
\end{table}

\begin{table}[t]
\centering
\small
\setlength{\tabcolsep}{4pt}     
\renewcommand{\arraystretch}{1.1}

\caption{
Predicted integrated reactor $\bar{\nu}_e$ flux $\Phi$ and IBD event rate for underground laboratories in 2030. The MSW columns adopt the three-flavor matter-enhanced neutrino-oscillation formalism, incorporating a three-dimensional Earth-density model and employing the Strang-splitting method as the baseline computational approach. 
$\Phi\,[\mathrm{cm^{-2}\,s^{-1}}]$ is integrated over $E_\nu\in[0.1,10]$ MeV.
$\Phi\,[\mathrm{cm^{-2}\,s^{-1}}]$ is integrated over $E_\nu\in[0.1,10]$ MeV. The definitions for
systematic errors $\Delta_{\rm \mathcal{M}}$, $\Delta_{\rm \oplus}$, and $\Delta_{\rm N/I}$ are given in the text.
}

\begin{tabular}{
l
c
c
c
c
c
c
c
}
\toprule
& \multicolumn{2}{c}{Vacuum} & \multicolumn{2}{c}{MSW} & \multicolumn{3}{c}{|Syst.|[\%]} \\
\cmidrule(lr){2-3}\cmidrule(lr){4-5}\cmidrule(lr){6-8}
Site &
{$\Phi$ [$\times10^{4}$]} &
{TNU} &
{$\Phi$ [$\times10^{4}$]} &
{TNU} &
{$\Delta_{\rm \mathcal{M}}$ } &
{$\Delta_{\rm \oplus}$ } &
{$\Delta_{\rm N/I}$ }\\
\midrule
ANDES        &$1.49^{+0.04}_{-0.04}$&$4.75^{+0.17}_{-0.16}$&$1.50^{+0.04}_{-0.04}$&$4.78^{+0.17}_{-0.17}$&$0.01$&$0.18$ &$0.02$       \\
Baksan       &$14.42^{+0.35}_{-0.42}$&$46.50^{+1.63}_{-1.61}$&$14.44^{+0.36}_{-0.41}$&$46.78^{+1.66}_{-1.60}$
&
$0.00$
&$0.31$&$0.01$\\
Boulby       &$251.64^{+8.20}_{-8.25}$&$1086.67^{+41.02}_{-40.36}$&$251.99^{+8.34}_{-8.09}$&$1087.78^{+40.71}_{-40.89}$ &
$0.00$
&$0.05$&$0.02$      \\
CallioLab    &$23.54^{+0.68}_{-0.58}$&$76.45^{+2.65}_{-2.66}$&$23.60^{+0.69}_{-0.57}$&$76.93^{+2.69}_{-2.64}$
&
$0.00$
&$0.35$ &
$0.00$
\\
CJPL         &$8.30^{+0.21}_{-0.23}$&$26.83^{+0.92}_{-0.93}$&$8.33^{+0.21}_{-0.24}$ &$27.00^{+0.93}_{-0.93}$&$0.02$&$0.21$ &$0.03$       \\
JUNO         &$676.23^{+19.84}_{-21.14}$&$1388.25^{+76.11}_{-70.53}$&$678.17^{+20.33}_{-20.70}$&$1401.25^{+76.35}_{-70.02}$&$0.02$& $0.52$ &$0.03$     \\
Kamioka      &$70.65^{+1.92}_{-1.91}$&$242.53^{+8.35}_{-8.34}$&$70.78^{+1.95}_{-1.88}$&$243.49^{+8.39}_{-8.31}$&$0.01$& $0.23$ &$0.02$      \\
LNGS         &$22.09^{+0.54}_{-0.63}$&$71.40^{+2.47}_{-2.46}$&$22.11^{+0.57}_{-0.61}$&$71.80^{+2.50}_{-2.45}$&$0.01$& $0.34$ &  $0.02$    \\
LSC &$67.62^{+1.63}_{-2.01}$&$221.29^{+7.65}_{-7.53}$&$67.75^{+1.66}_{-1.97}$&$222.35^{+7.73}_{-7.49}$&$0.01$&$0.31$& $0.02$      \\
LSM   &$138.34^{+3.45}_{-3.93}$&$476.94^{+16.18}_{-16.60}$&$138.64^{+3.49}_{-3.88}$&$479.61^{+16.36}_{-16.51}$&$0.01$&$0.35$  &$0.02$     \\
SNO          &$55.98^{+1.61}_{-1.37}$&$175.51^{+6.23}_{-6.08}$&$56.09^{+1.63}_{-1.35}$&$176.53^{+6.31}_{-6.03}$&$0.01$&$0.36$ & $0.02$     \\
SUPL         &$0.85^{+0.03}_{-0.02}$&$2.76^{+0.10}_{-0.10}$&$0.86^{+0.02}_{-0.02}$&$2.78^{+0.10}_{-0.10}$&
$0.00$
&$0.02$ & 
$0.01$
\\
SURF         &$9.13^{+0.25}_{-0.24}$&$29.43^{+1.02}_{-1.01}$&$9.16^{+0.24}_{-0.24}$&$29.64^{+1.03}_{-1.01}$&$0.03$& $0.25$ &$0.04$      \\
Yemilab      &$436.86^{+11.16}_{-13.26}$&$922.07^{+43.97}_{-42.26}$&$437.79^{+11.42}_{-12.94}$&$931.85^{+45.43}_{-41.14}$&$0.01$&$0.51$ &$0.03$       \\
\bottomrule
\end{tabular}

\label{table:prediction_2030}
\end{table}

%% file: Appendix.tex
\renewcommand{\appendixname}{Appendix}

\appendix
\section*{Appendix}
\renewcommand{\thesubsection}{A.\arabic{subsection}}
\renewcommand{\theequation}{A.\arabic{equation}}
\renewcommand{\thefigure}{A.\arabic{figure}}
\renewcommand{\tablename}{Table}
\renewcommand{\thetable}{A.\arabic{table}}
\setcounter{table}{0}
\setcounter{equation}{0}
\setcounter{subsection}{0}

\subsection{First-order Approximation Approach to the Reactor Neutrino MSW Effects}
\label{sec:1st_order}
In this section, we will evaluate Eq.~\eqref{eq:layer_product} in the flavor basis, where $H_{f,k}=H_\text{vac}(E)+H_\text{mat}(x_k)$, with 
 \begin{align}
 H_\text{vac}= U^*\mathrm{diag}\left (0,\ \frac{\Delta m_{21}^2}{2E},\ \frac{\Delta m_{31}^2}{2E}\right )U^T , 
 \quad
 H_{\text{mat},k}=\mathrm{diag}(V^{CC}_k,\ 0,\ 0),
 \end{align}
where $U$ is the PMNS matrix and $V^{CC}_k\equiv \eta\ \sqrt{2}G_F N_e(x_k)$ denotes the charged-current matter potential evaluated at the midpoint of layer step $k$, and $\eta=+1$ for neutrinos and $\eta=-1$ for antineutrinos. Furthermore, we use $V_{\mathrm{eff}}=c^2_{13}V^{CC}$, which is the matter potential projected onto the effective $(1$--$2)$ subsystem.

Within each layer step, the electron density is assumed to be constant, so the Hamiltonian
 $H_{f,k}$ is $x$-independent. Therefore, the evolution operator
across the $k$-th layer step is exactly
\begin{equation}
S_{f,k}\equiv S_{f}(x_{k},x_{k-1})=\exp\!\left[-i\,H_{f,k}\,\Delta x_k\right].
\end{equation}
The ordered product gives the total evolution operator
$S_{f}(L,0)\approx \prod_{k=N}^{1} S_{f,k}$.
For each layer step, the Hamiltonian can be diagonalized as
\begin{equation}
H_{f,k}=U_k^{m*}\,\Lambda_k^{\rm full}\,U_k^{mT},\qquad
\Lambda_k^{\rm full}=\mathrm{diag}(\lambda_{1,k},\lambda_{2,k},\lambda_{3,k}).
\end{equation}
Since adding $\alpha I$ to $H_{f,k}$ only produces a global (unphysical) phase, we subtract $\lambda_{1,k}I$ and work with
\begin{equation}
\tilde H_{f,k}=H_{f,k}-\lambda_{1,k}I
=U_k^{m*}\,\Lambda_k\,U_k^{mT},\qquad
\Lambda_k=\mathrm{diag}(0,\lambda^{21}_{k},\lambda^{31}_{k}),
\end{equation}
where $\lambda^{21}_{k}\equiv \lambda_{2,k}-\lambda_{1,k}$ and $\lambda^{31}_{k}\equiv \lambda_{3,k}-\lambda_{1,k}$. (Equivalently, in the $2\times2$ (1--2) block one may use
$\Lambda_k=\mathrm{diag}(\frac{-\lambda^{21}_{k}}{2},\ \frac{+\lambda^{21}_{k}}{2})$,
which differs from $\mathrm{diag}(0,\lambda^{21}_{k})$ only by an overall phase.) Therefore, Eq.~\eqref{eq:layer_product} can be written as,
	\begin{align}
	S_{f}(x_N=L,0) \;& \approx\; \prod_{k=N}^{1} U_k^{m*}\,D_k\,U_k^{mT}, \nonumber\\
	& = U_{N}^{m*}\,D_{N}\,U_{N}^{m\ T}\cdot U_{N-1}^{m\ *}\,D_{N-1}\,U_{N-1}^{m\ T}\cdots U_{1}^{m\ *}\,D_1\,U_{1}^{m\ T}, \nonumber\\
    & = U_{N}^{m*} [\prod_{k=N}^{2}(D_k U_{k, k-1})]D_1U_{1}^{m\ T},
	\label{eq:layer_product_2}
	\end{align} 
where 
\begin{align}
D_k &\equiv  \mathrm{exp}(-i\Lambda_k \Delta x_k)\\
&=\begin{cases}
\mathrm{diag}(1,\ e^{-i\phi^{21}_k},\ e^{-i\phi^{31}_k}), &
 \text{in}~SU(3) \\[6pt]
\mathrm{diag}(\ e^{-i\phi^{21}_k/2},\ e^{i{\phi^{21}_k/2}}),& 
 \text{in}~SU(2) \text{(1--2) block}
\end{cases}
\end{align}
\begin{align}
\phi^{21}_k \equiv \lambda^{21}_k \Delta x_k, \quad \phi^{31}_k \equiv \lambda^{31}_k \Delta x_k,
\end{align}
and $U_{k, k-1}(-\Delta\theta_{k-1})=U_{k}^{m\ T} U_{k-1}^{m\ *}$ can be considered as a rotation between layers, which can be written in the lowest order of rotation~\cite{Ioannisian:2017dkx} as
\begin{align}
U_{k, k-1}=I-i \sigma_2 \sin(\Delta \theta_{k-1}),
\label{eq:su2_form}
\end{align}
in $SU(2)$, where $\sigma_2=\begin{pmatrix}
0 & -i\\
i & 0 \\
 \end{pmatrix}$ is Pauli matrix;
\begin{align}
U_{k, k-1}=I-i \lambda_2 \sin(\Delta \theta_{k-1})
\label{eq:su3_form}
\end{align}
in $SU(3)$, where $\lambda_2$ is Gell-Mann matrix:
\begin{equation}
\lambda_{2}=\begin{pmatrix}
0 & -i & 0\\
i & 0 & 0 \\
0 & 0 & 0
 \end{pmatrix}.
\end{equation}
In Eq.~\eqref{eq:layer_product_2}, absorb the $\cdots D_1U_{1}^{m\ T}$ into the whole multiplicative expression by adding $U_0^*U_0^T=U_{\rm{vac}}^*U_{\rm{vac}}^T$ on the right-hand side, and apply Eq.~\eqref{eq:su2_form} or Eq.~\eqref{eq:su3_form} as an approximation, which gives,
\begin{align}
&\quad \quad S_f(x_N=L,0)
\approx U_{N}^{m*}\!\Big[\prod_{k=N}^{1}(D_k\,U_{k,k-1})\Big]\;U_{\rm vac}^{T} \nonumber\\[9pt]
&=\begin{cases}
U_{N}^{m*}\prod_{k=N}^{1}\ D_k[I-i\sigma_2\sin(\Delta\theta_{k-1})]U_{\rm vac}^{T}, & \text{in }SU(2) \text{ block} \\[9pt]
U_{N}^{m*}\prod_{k=N}^{1}\ D_k[I-i \lambda_2 \sin(\Delta \theta_{k-1})]U_{\rm vac}^{T}, & \text{in }SU(3).
\end{cases}
\label{eq:layer_product_3}
\end{align}
In terms of first order approximation expanded in $SU(2)$, by assuming that the last layer is in vacuum, the multiplicative expression can be evaluated in $2 \times 2$ blocks first and kept only up to $\Delta \theta_k$ as,
\begin{align}
U_\text{vac}^{*}\left [D(\phi_{21}^{tot})-i\sum_{j=1}^{N}\sin(\Delta\theta_{j})D(\phi_{21}^{a})\sigma_2 D(\phi_{21}^{b})\right]U_{\rm vac}^{T},
\label{eq:fo_sum_su2}
\end{align}
where $\phi^{21}_\text{tot}$ is the total phase acquired in the Earth,
\begin{align}
\phi^\text{tot}_{21}=\sum_{k=1}^{N}\phi^{21}_k,
\end{align}
and with total phases acquired before or after jump $j$~\cite{Ioannisian:2017dkx},
\begin{align}
\phi_{21}^{b}=\sum_{k=1}^{j}\phi_{21,k}, \quad
\phi_{21}^{a}=\sum_{k=j+1}^{N}\phi_{21,k}.
\end{align}
The first and second terms of Eq.~\eqref{eq:fo_sum_su2} are denoted as
\begin{align}
&A^{\mathrm{2fl},0} = U_\text{vac}^{*}\ D(\phi_{21}^\text{tot})U_{\rm vac}^{T} \nonumber\\
&A^{\mathrm{2fl},1} = -i U_\text{vac}^{*}\sum_{j=1}^{N}\sin(\Delta\theta_{j})D(\phi_{21}^{a})\sigma_2 D(\phi_{21}^{b})\ U_{\rm vac}^{T}
\end{align}
Therefore,  the survival probability of antineutrino, $P^\mathrm{2fl}_{ee}\approx|\langle \bar{\nu}_e|S_f|\bar{\nu}_e \rangle|^2=|S_{fee}|^2 $. Neglecting the CP-violation phase and keeping only up to $\Delta \theta_k$, the survival probability is written as
\begin{align}
&P^{2\mathrm{fl}}_{ee}
\simeq |A^{\mathrm{2fl},0}_{ee}|^2+2\Re\!\left[(A^{\mathrm{2fl},0}_{ee})^*\cdot A^{\mathrm{2fl},1}_{ee}\right]\nonumber\\
&=1-\sin^2(2\theta_{12})\sin^2\!\left(\frac{\phi^{\mathrm{tot}}_{21}}{2}\right) \nonumber\\
&\quad +2\sin(2\theta_{12})\cos(2\theta_{12})\sin\!\left(\frac{\phi^{\mathrm{tot}}_{21}}{2}\right)
\sum_{j=1}^{N}\sin(\Delta \theta_j)\,
\sin\!\left(\frac{\phi^{\mathrm{tot}}_{21}}{2}-\phi^{a}_{21,j}\right).
\label{eq:Pee_2fl}
\end{align}
Using integrals by parts and setting $sin(\Delta \theta_j) \rightarrow \Delta \theta_j$, the above equation can be rewritten as an integral form in terms of $x$,
\begin{align}
&P^{2\mathrm{fl}}_{ee}
\simeq |A^{\mathrm{2fl},0}_{ee}|^2+2\Re\!\left[(A^{\mathrm{2fl},0}_{ee})^*\cdot A^{\mathrm{2fl},1}_{ee}\right]\nonumber\\
&=1-\sin^2(2\theta_{12})\sin^2\!\left(\frac{\phi^{\mathrm{tot}}_{21}}{2}\right) \nonumber\\
&\quad -\sin^2(2\theta_{12})\cos(2\theta_{12})\sin\!\left(\frac{\phi^{\mathrm{tot}}_{21}}{2}\right)
\int^{L}_{0}V_{\mathrm{eff}}(x)\,
\cos\!\left[\frac{\phi^{\mathrm{tot}}_{21}}{2}-\phi^{m}_{x \rightarrow L}(x)\right]dx,
\label{eq:Pee_2fl_int}
\end{align}
where
\begin{align}
\phi^{m}_{x \rightarrow L}(x)=\int^{L}_{x}{dx' \frac{\Delta m^2_{21}}{2E}\sqrt{\left [\cos(2\theta_{12})-\frac{2V_{\mathrm{eff}}(x')E}{\Delta m^2_{21}}\right ]^2+\sin^2(2\theta_{12})}},
\label{eq:Phi_x_L}
\end{align}
and
\begin{align}
\phi^\text{tot}_{21}(x)=\int^{L}_{0}{dx \frac{\Delta m^2_{21}}{2E}\sqrt{\left [ \cos(2\theta_{12})-\frac{2V_{\mathrm{eff}}(x)E}{\Delta m^2_{21}}\right ]^2+\sin^2(2\theta_{12})}}.
\label{eq:phi_tot}
\end{align}
We employ two first-order treatments of the Earth-matter (MSW) effect.
In the effective $SU(2)$ $(1$--$2)$ description, the third state is a spectator and the rapid
$1$--$3$ oscillations are averaged (i.e.\ effectively decoupled), giving
\begin{equation}
P^{3\mathrm{fl}}_{ee}\simeq c_{13}^4\,P^{2\mathrm{fl}}_{ee}+s_{13}^4 .
\end{equation}
where $c_{13}\equiv \cos(\theta_{13})$ and $s_{13}\equiv \sin(\theta_{13})$. The $SU(3)$ approach keeps the full three-flavor structure of the layer-to-layer rotations
at the same perturbative order in $\Delta\theta_j$, and therefore retains the interference
terms involving the $1$--$3$ phase (e.g.\ $\Delta m_{31}^2 L/2E$) in both $|A^0_{ee}|^2$
and the first-order correction. The $SU(2)$ result is recovered from the $SU(3)$ expression after averaging over the
unresolved $1$--$3$ oscillations, as expected for finite energy resolution and/or baseline smearing, 
where $\sin^2(\Delta m_{31}^2 L/4E)$ is effectively replaced by its averaged value.
Applying the same calculation procedure to the $SU(3)$ expansion above, we obtain the survival probability as
\begin{align}
|A^{\mathrm{3fl},0}_{ee}|^2 =1 &-\sin^2(2\theta_{12})c_{13}^4\sin^2\left (\frac{\phi^\text{tot}_{21}}{2}\right ) \nonumber\\
&-\sin^2(2\theta_{13})c_{12}^2\sin^2\left (\frac{\Delta m^2_{31}L}{4E}\right ) \nonumber\\
&-\sin^2(2\theta_{13})s_{12}^2\sin^2\left (\frac{\Delta m^2_{31}L}{4E}-\frac{\phi^\text{tot}_{21}}{2}\right ),
\label{eq:A_3fl_0}
\end{align}
with
\begin{align}
2\,\Re[&(A^{\mathrm{3fl},0}_{ee})^* A^{\mathrm{3fl},1}_{ee}]
= -\,sin^2(2\theta_{12})c_{13}^{2} \nonumber\\
&\quad \times
\Bigg[
c_{13}^{2}\cos(2\theta_{12})\,\sin\!\left(\frac{\phi^\text{tot}_{21}}{2}\right)
+ s_{13}^{2}\,\sin\!\left(\frac{\phi^\text{tot}_{21}}{2}-\frac{\Delta m^{2}_{31}L}{2E}\right)
\Bigg] \nonumber\\
&\quad \times \int_{0}^{L}V_{\mathrm{eff}}(x)\,
\cos\!\left[\frac{\phi^\text{tot}_{21}}{2}-\phi^m_{x \rightarrow L}(x)\right]dx.
\label{eq:A_3fl_1}
\end{align}
where $c_{12}\equiv \cos(\theta_{12})$ and $s_{12}\equiv \sin(\theta_{12})$. The survival probability then is,
\begin{equation}
P^\mathrm{3fl}_{ee}\simeq|A^{\mathrm{3fl},0}_{ee}|^2+2\,\Re[(A^{\mathrm{3fl},0}_{ee})^* \cdot A^{\mathrm{3fl},1}_{ee}],
\label{eq:3fl_Pee}
\end{equation}
note that Eqs.~\eqref{eq:A_3fl_0} and~\eqref{eq:A_3fl_1} reduce to Eq.~\eqref{eq:Pee_2fl} while setting $c_{13} \rightarrow 1$ and $s_{13} \rightarrow 0$.

\subsection{Cross-check with other studies}

To validate our computational results, we cross-reference a selection of them with findings reported in the existing literature.

\subsubsection{Cross-check of total IBD rate of multiple site}
\begin{table}[t]
  \centering
  \small
  \setlength{\tabcolsep}{8pt}
  \renewcommand{\arraystretch}{1.15}

  \begin{tabular*}{\textwidth}{@{\extracolsep{\fill}} l
    c
    c
    c}
    \toprule
    Site &
    {$\mathrm{IBD\ Signal}\pm {\mathrm{spread}}$ [TNU]} &
    Ref. [TNU] &
    {$\Delta$ [\%]} \\
    \midrule
    JUNO(<~300~km) & $1351.6^{+40.1}_{-40.3}$ & 1336.2 & +1.2 \\
    JUNO(>~300~km) & $41.5^{+1.3}_{-1.3} $  & 42.2 &  -1.7  \\
    SNO+ & $170.5_{-5.9}^{+6.1}$    &  $174.1^{+7.5}_{-7.5}$(converted)   &    -1.8   \\
    \bottomrule
  \end{tabular*}

  \caption{Cross-check of reactor $\bar{\nu}_e$ predictions under vacuum oscillations.
  The IBD rate over $E_\nu>1.806$~MeV. And the $\bar{\nu}_e$ energy spectrum of the SM2023 model is applied}
  \label{tab:xc_prev_sm}
\end{table}
In this section, we will compare our calculated event rates (in TNU) with results from other analyses(JUNO~\cite{JUNO:2025}, SNO~\cite{SNO:2025ktonne}). For consistency, all calculations in this cross-check are performed under the assumption of vacuum oscillations. 
The literature comparisons in this section are performed according to a unified prescription. If a parameter is explicitly given in the reference under consideration, we directly adopt the published value. If a required input is not explicitly specified, we instead use the default configuration described in ~\ref{sec:default-setup}. In such cases, the corresponding default uncertainties are propagated through Monte Carlo sampling to obtain the spread quoted in this work. Accordingly, the central values are intended to reproduce the literature as closely as possible. In contrast, the spreads should be understood as uncertainty estimates derived from our unified completion scheme rather than as exact reproductions of the original literature error budget.

For the nearby reactors within 300~km, the JUNO Collaboration estimated the reactor $\bar{\nu}_e$ background using the Huber--Mueller model, weighted by the IBD cross section, and corrected using the Daya Bay observation. The fission fractions were fixed to 0.58, 0.07, 0.30, and 0.05 for $^{235}\mathrm{U}$, $^{238}\mathrm{U}$, $^{239}\mathrm{Pu}$, and $^{241}\mathrm{Pu}$, respectively, and corrections for non-equilibrium effects and spent nuclear fuel were included. The quoted rate of 43.2 events/day ($1336.2$~TNU) assumes an overall IBD selection efficiency of 82.2\% and an average reactor duty cycle(load factor) of 11/12. For reactors beyond 300~km, JUNO updated the world-reactor contribution using the IAEA PRIS 2024 database and reported a reference value of $42.2$~TNU. The nominal oscillation parameters adopted in the spectral analysis were taken from PDG 2024~\cite{JUNO:2025}. In the present work, rather than attempting to fully reproduce the JUNO implementation of the HM+$\,$Daya Bay--corrected reactor spectrum, we adopt the SM2023 model as the nominal spectral input. Therefore, the quoted spread reflects only the uncertainty associated with the adopted spectral model. As shown in Table~\ref{tab:xc_prev_sm}, the JUNO reference values are covered by our estimated uncertainty bands.

For comparison with SNO+, we use a converted reference value derived from Table II of Ref.~\cite{SNO:2025ktonne}, where the expected reactor IBD count is reported as $140\pm6$, together with expected geoneutrino counts of 29 from the $^{238}$U chain and 8 from the $^{232}$Th chain. In the same paper, the SNO+ collaboration states that their geoneutrino model predicts $36.3\pm8.7$ TNU and $9.7\pm2.3$ TNU from $^{238}$U and $^{232}$Th, respectively, corresponding to about 37 expected geo-$\bar {\nu}_e$ IBD events in the full dataset. We therefore infer a count-to-TNU conversion factor of $(36.3+9.7)/(29+8)=46.0/37$, and convert the reactor expectation into a reference value of $174.1\pm7.5$ TNU. Here, the quoted uncertainty is obtained by propagating only the literature uncertainty on the reactor IBD count. For the comparison setup, Table II of Ref.~\cite{SNO:2025ktonne} states only that the expected counts use oscillation parameters from the PDG 2025 update. We therefore adopt our default PDG-2025 normal-ordering values as the nominal input for oscillations. Since the SNO+ paper does not provide a complete numerical specification of the fission fractions $f_i$ or the per-fission energies $Q_i$, we retain our default setup for these inputs. The SNO+ analysis uses data acquired from May 2022 through July 2025, with reactor thermal powers tracked daily using public electrical output data from IESO in Ontario and the NRC in the USA, cross-checked against monthly IAEA averages. For our comparison, we instead use the 2024 operating fleet reported in IAEA/OPEX/2025 as a representative full-year proxy within that interval. As shown in Table~\ref{tab:xc_prev_sm}, the SNO+ reference value is covered by our estimated uncertainty bands.

\subsubsection{Probability-level MSW benchmark against Ref.~\cite{JUNO:2016}}
\begin{figure}[t]
    \centering
    \includegraphics[width=0.9\linewidth]{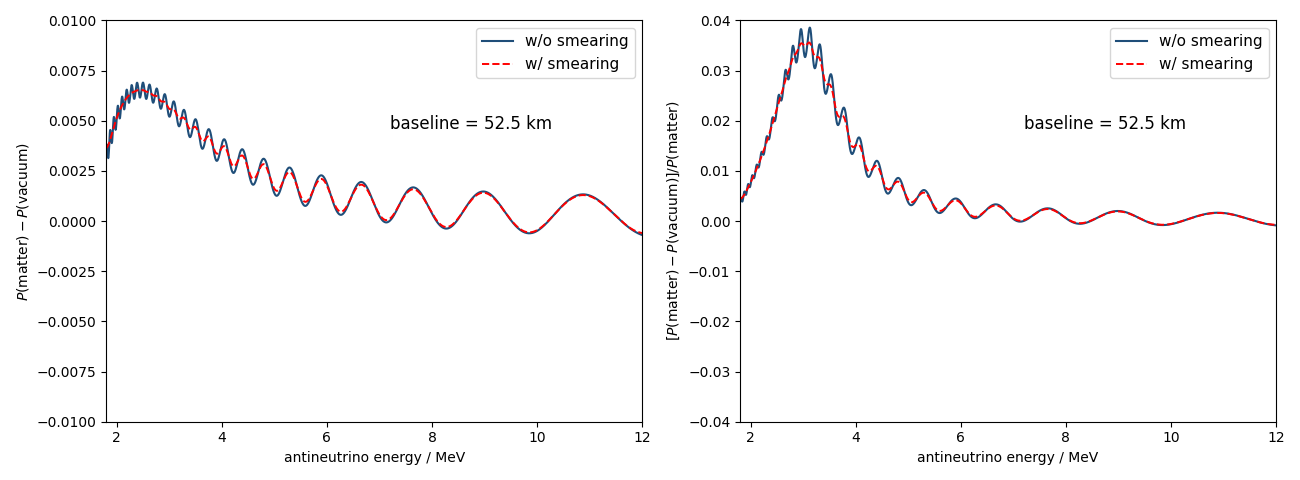}
    \caption{Reproduction of the probability-level benchmark in Fig.~1 of Ref.~\cite{JUNO:2016}. Shown are the absolute (left) and relative (right) differences between the matter-corrected and vacuum antineutrino survival probabilities for a constant-density setup with $L=52.5$~km and $\rho=2.6~\mathrm{g/cm^3}$. Solid curves correspond to the true antineutrino energy, while dashed curves include a Gaussian energy smearing of $3\%/\sqrt{E(\mathrm{MeV})}$. The results are obtained with our Strang-splitting solver using the parameter setup of Ref.~\cite{JUNO:2016}.}
    \label{fig:JUNO_compare}
\end{figure}

To further validate the probability-level implementation of the MSW evolution, we reproduce Fig.~1 of Ref.~\cite{JUNO:2016} using the same constant-density benchmark setup. This benchmark corresponds to a medium-baseline reactor antineutrino oscillation configuration with $L=52.5$~km, $\rho=2.6~\mathrm{g/cm^3}$, and a Gaussian energy resolution of $3\%/\sqrt{E(\mathrm{MeV})}$. As shown in Fig.~\ref{fig:JUNO_compare}, our Strang-splitting calculation reproduces the published absolute and relative matter--vacuum probability differences, both before and after energy smearing. 

\subsection{Convergence Test of MSW Calculation Methods}
\label{sec:converge_test}

\begin{figure}[!htbp]
    \centering
    \includegraphics[width=1.0\linewidth]{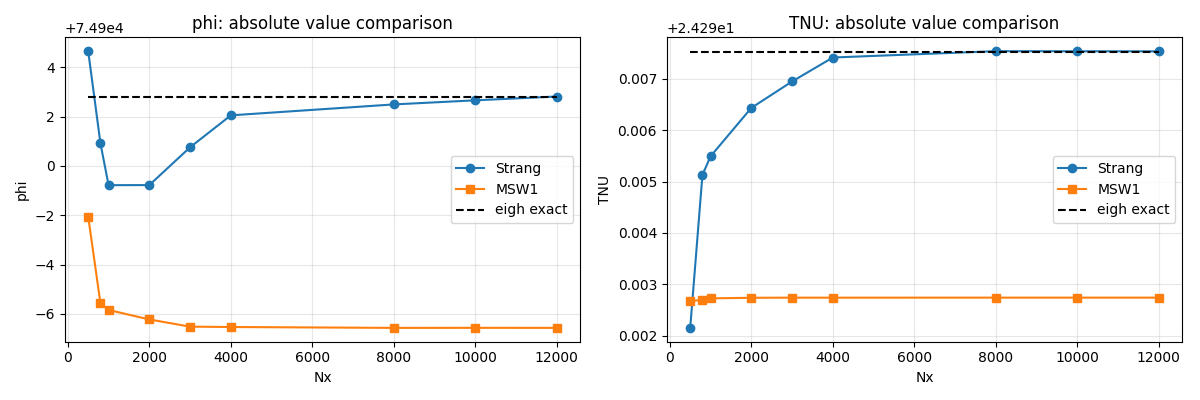}
    \caption{Convergence test of the MSW solvers for CJPL. Left panel: integrated reactor $\bar{\nu}_e$ flux $\Phi$. Right panel: IBD event rate in TNU. The blue and orange curves show the results obtained with the Strang-splitting solver in Eq.~\eqref{eq:strang_P} and the first-order approximation in Eq.~\eqref{eq:S_probibility}, respectively. The dashed line denotes the reference solution obtained by directly diagonalizing the layer Hamiltonians $H_i$ in Eq.~\eqref{eq:layer_product} with \texttt{linalg.eigh}, and $Nx = 12000$.}
    \label{fig:converge_test}
\end{figure}
Figure~\ref{fig:converge_test} shows that the Strang-splitting result approaches the reference solution as the number of path segments $N_x$ increases. For the TNU observable, the relative deviation of the Strang result decreases from $2.2\times10^{-4}$ at $N_x=500$ to $4.7\times10^{-6}$ at $N_x=4000$, and further to $3.5\times10^{-7}$ at $N_x=8000$. For the integrated flux $\Phi$, the convergence is somewhat slower and exhibits mild non-monotonic behavior at small $N_x$, but the relative deviation is still reduced to $1.0\times10^{-5}$ at $N_x=4000$ and $4.2\times10^{-6}$ at $N_x=8000$. At $N_x=8000$, the Strang result gives $\Phi=74902.514$ compared with the reference value $\Phi=74902.827$, and $24.297531$ TNU compared with the reference value $24.297523$ TNU. We therefore adopt $N_x=8000$ in the production calculation. At the precision relevant for this work, the Strang result is numerically indistinguishable from the reference solution. By contrast, the first-order approximation quickly stabilizes but remains systematically offset from the reference, at the level of $\sim1.3\times10^{-4}$ for $\Phi$ and $\sim2.0\times10^{-4}$ for TNU. This result confirms that the first-order treatment is adequate as a diagnostic approximation, while the Strang-splitting solver is preferred as the baseline MSW treatment in the main text.

A subtle caveat is that, while the Strang-splitting solver converges very well for integrated observables such as $\Phi$ and TNU, the low energy points can exhibit a numerical step-size resonance when the fast $1$--$3$ vacuum phase accumulated within a single layer satisfies $\Delta m_{31}^2 \Delta x /(4E_\nu) \approx n\pi$; this is a discretization artifact associated with coherent accumulation of the leading $O(\Delta x^3)$ splitting error, and for the terrestrial trajectories considered here with $N_x=8000$, the first such resonance would occur only at $E_\nu \lesssim 1.6~\mathrm{MeV}$ even for an Earth-diameter chord, so it has no practical impact on the quoted integrated $\Phi$ and TNU results.